\documentclass[oldversion]{aa}
\usepackage{graphicx}
\usepackage{hyperref}

\usepackage{arabtex}
\usepackage{utf8}
\setcode{utf8}

\usepackage{CJK}

\def\TKIN {$T_{\rm kin}$}

\def\H0 {$H_{\rm o}$}

\def\cmsq  {$\hbox{{\rm cm}}^{-2}$}

\def\MOLH {\hbox{${\rm H}_2$}}                    
                     
\def\AMM {\hbox{${\rm NH}_{3}$}}

\def\CH3C2H {\hbox{${\rm CH}_3{\rm C}_2{\rm H}$}}

\def\arcsec {\hbox{$^{\prime\prime}$}}

\def\ffas {\hbox{$\,.\!\!^{\prime\prime}$}}

\def\ffs {\hbox{$\,.\!\!^{\rm s}$}}

\def \ga{\mathrel{\mathchoice   {\vcenter{\offinterlineskip\halign{\hfil
$\displaystyle##$\hfil\cr>\cr\sim\cr}}}
{\vcenter{\offinterlineskip\halign{\hfil$\textstyle##$\hfil\cr
>\cr\sim\cr}}}
{\vcenter{\offinterlineskip\halign{\hfil$\scriptstyle##$\hfil\cr
>\cr\sim\cr}}}
{\vcenter{\offinterlineskip\halign{\hfil$\scriptscriptstyle##$\hfil\cr
>\cr\sim\cr}}}}}

\begin{document}

\title{Ammonia in the hot core W51-IRS2: Maser line profiles, variability
       and saturation}

\author{E. Alkhuja  (\<عِمَاد الخُجَا>) \inst{1,2},
        C. Henkel \inst{1,3},
         {\protect\begin{CJK*}{UTF8}{gkai}Y. T. Yan (闫耀庭)\protect\end{CJK*}}     \inst{1},
        B. Winkel \inst{1},
        Y. Gong   \inst{4,1},
        G. Wu     \inst{1},
        T.L. Wilson \inst{1},
        A. Wootten \inst{5},
        A. Malawi (\<عبدالرحمن ملاوي>)  \inst{2}
       }

\offprints{F. Author, \email{E. Alkhuja}}

\institute{Max-Planck-Institut f{\"u}r Radioastronomie,
           Auf dem H{\"u}gel 69, 53121 Bonn, Germany
 \and      Astronomy Department, Faculty of Science,
           King Abdulaziz University, P.O. Box 80203, 
           Jeddah 21589, Saudi Arabia
 \and      Xinjiang Astronomical Observatory, Chinese
           Academy of Sciences, 830011 Urumqi, PR China 
 \and      Purple Mountain Observatory (PMO) and Key Laboratory
           of Radio Astronomy, Chinese Academy of Sciences,
           10 Yuanhua Road, Nanjing 210023, PR China  
 \and      National Radio Astronomy Obsrevatory, 520 
           Edgemont Road, Charlottesville, VA 22903-2475, 
           USA
  }

\date{Received 29 March 2025 / Accepted 19 June 2025}
 
\abstract
{
W51-IRS2 is known to be one of the most prolific sources of interstellar ammonia (NH$_3$) 
maser lines. So far, however, many of these inversion lines have rarely been studied. 
Here we report spectrally resolved line profiles for the majority of detected features 
and provide information on the variability of these maser components between 2012 and 2023. 
This includes the first tentative detection of a ($J$,$K$) = (5,2) maser in the interstellar 
medium and the first tentative detection of a (6,4) maser in W51-IRS2. Furthermore, we report 
for the first time NH$_3$ (9,6) maser emission below Local Standard of Rest velocities of 
50\,km\,s$^{-1}$ in this source as well as double maser features occasionally seen in 
other transitions. The detected maser lines strongly indicate vibrational 
pumping by $\approx$10\,$\mu$m photons, which must be abundant due to the high kinetic 
temperature ($\approx$300\,K) of the ammonia emitting gas. The detection of vibrationally 
excited NH$_3$, suggesting a vibrational excitation temperature consistent with the kinetic
one,  and a comparison with measured SiO line profiles is also presented. For the
(10,7) line, we find a tentative correlation between flux density and line width 
compatible with unsaturated maser emission. The velocity drift of the so-called 
45\,km\,s$^{-1}$ maser features, reported to be +0.2\,km\,s$^{-1}$\,yr$^{-1}$ between 
1996 and 2012, has either slowed down to values $<$0.1\,km\,s$^{-1}$ or has entirely 
disappeared. In 2023, the component is only seen in ammonia inversion lines that are 
located at least 800\,K above the ground state. The other features have faded. Possible
scenarios explaining this phenomenon are discussed. }


\titlerunning{NH$_3$ masers in W51-IRS2}

\authorrunning{Authors}

\maketitle

\section{Introduction}

Ammonia (\AMM ) provides unique opportunities to trace molecular cloud
excitation up to temperatures of $\ga$1000\,K (e.g., Wilson et al. 2006)
by observing its inversion transitions within a very limited frequency
interval (20 -- 45\,GHz). Dozens of inversion lines can be detected in
this frequency range, provided kinetic temperatures and column densities
are high enough. Warmer conditions are characteristic for ``hot cores'',
dense molecular condensations near sites of massive star formation that are
chemically affected by dust grain mantle evaporation through sudden irradiation 
by an intense UV radiation field from short-lived OB stars (e.g., Genzel
et al. 1982; Henkel et al. 1987; Walmsley et al. 1987; Brown et al. 1988;
Charnley et al. 1992; Caselli et al. 1993; Wakelam et al. 2005; Bisschop
et al. 2007; Garrod et al. 2013). In such environments large amounts of hydrogenated 
molecules may enter the gas phase, with ammonia and water vapor among the most 
notable species. The warm dense clumps are characterized by temperatures 
\TKIN $>$ 100\,K, $X$(\AMM ) = $N$(\AMM )/$N$(\MOLH ) $\approx$ 10$^{-5...-6}$ 
and source averaged ammonia column densities in excess of 10$^{18}$\,\cmsq 
(e.g., Mauersberger et al. 1987).

Not all of the ammonia inversion lines are thermally excited. ($J,K$) = (3,3)
maser emission was first detected by Wilson et al. (1982) towards the star
forming region W\,33. This maser is readily explained by the fact that the two
lowest $K$-ladders of the symmetric ortho-NH$_3$ molecule, those for
$K$ = 0 and 3, are radiatively connected and that the $K$=0 ladder is the one
devoid of inversion doublets (it is characterized instead by single rotational
states). The single (0,0) state, the important and usually well populated
ground state, is collisionally connected to only the upper level of the 
(3,3) inversion doublet, which inverts the populations of the two (3,3) 
levels under specific physical conditions (e.g., Walmsley \& Ungerechts 1983). 

By 18 years ago, several maser lines were already identified (e.g., Guilloteau 
et al. 1983; Gaume et al. 1996; Madden et al. 1986; Mauersberger et al. 1986a, 
1987, 1988; Cesaroni et al. 1992; Mangum \& Wootten 1994; Beuther et al. 2007), 
with detections not only including the (3,3) line of $^{14}$NH$_3$ but 
also that of $^{15}$NH$_3$ (Mauersberger et al. 1986b; Schilke et al. 1991). 
Other inversion doublets were also found to be masing. The numerous lines 
detected during the last decade (e.g. Henkel et al. 2013; Mills et al. 2018; 
Mei et al. 2020; Yan et al. 2022a,b, 2024), first towards W51 and then also towards 
Sgr\,B2 and other massive star forming regions, often include lines of 
ortho-NH$_3$ ($K$ = 0, 3, 6, 9, ...). However, many transitions of 
para-NH$_3$ ($K$ = 1, 2, 4, 5, 7, 8, ...) were also found. To date, maser 
transitions were detected in more than 30 different ammonia inversion lines 
(for a list, see table~A.1 in Yan et al. 2024). These are usually  
found at levels hundreds of Kelvins above the ground state, with the ($J$,$K$) 
= (12,12) transition at $\approx$1450\,K reaching the highest value. With 
the exception of the (1,1) (Gaume et al. 1996), (2,2) (Mills et al. 2018), 
(3,3) (Wilson et al. 1982), (5,5) (Cesaroni et al. 1992), (6,6) (Beuther 
et al. 2007), (7,7) (9,9) and (12,12) (Henkel et al. 2013) inversion doublets, 
the maser lines originate from non-metastable levels, requiring extremely 
high densities ($\ga$10$^8$\,cm$^{-3}$) and/or intense infrared radiation 
fields to become populated. While the (3,3) maser lines are readily reproduced 
by large velocity gradient (LVG) radiative transfer models, explanations for 
the other masers are difficult to attain. These possibly require a quantitative 
understanding of vibrational excitation, which is a problem, because until 
recently collisional rates were only available up to the (6,6) state of the 
ground vibrational level and because IR line overlaps with transitions from 
other molecular species might also play a role. Quite recently, Demes et al. 
(2023) published collision rates for ortho-NH$_3$ up to ($J$,$K$) = (10,9), 
including the (9,6) inversion doublet and para NH$_3$ up to (10,10), but 
without the (10,7) transition, while Loreau et al. (2023) reported collision 
rates for all hyperfine levels up to $J$ = 4.

\begin{figure}[t]
\vspace{0.0cm}
\resizebox{9.0cm}{!}{\includegraphics{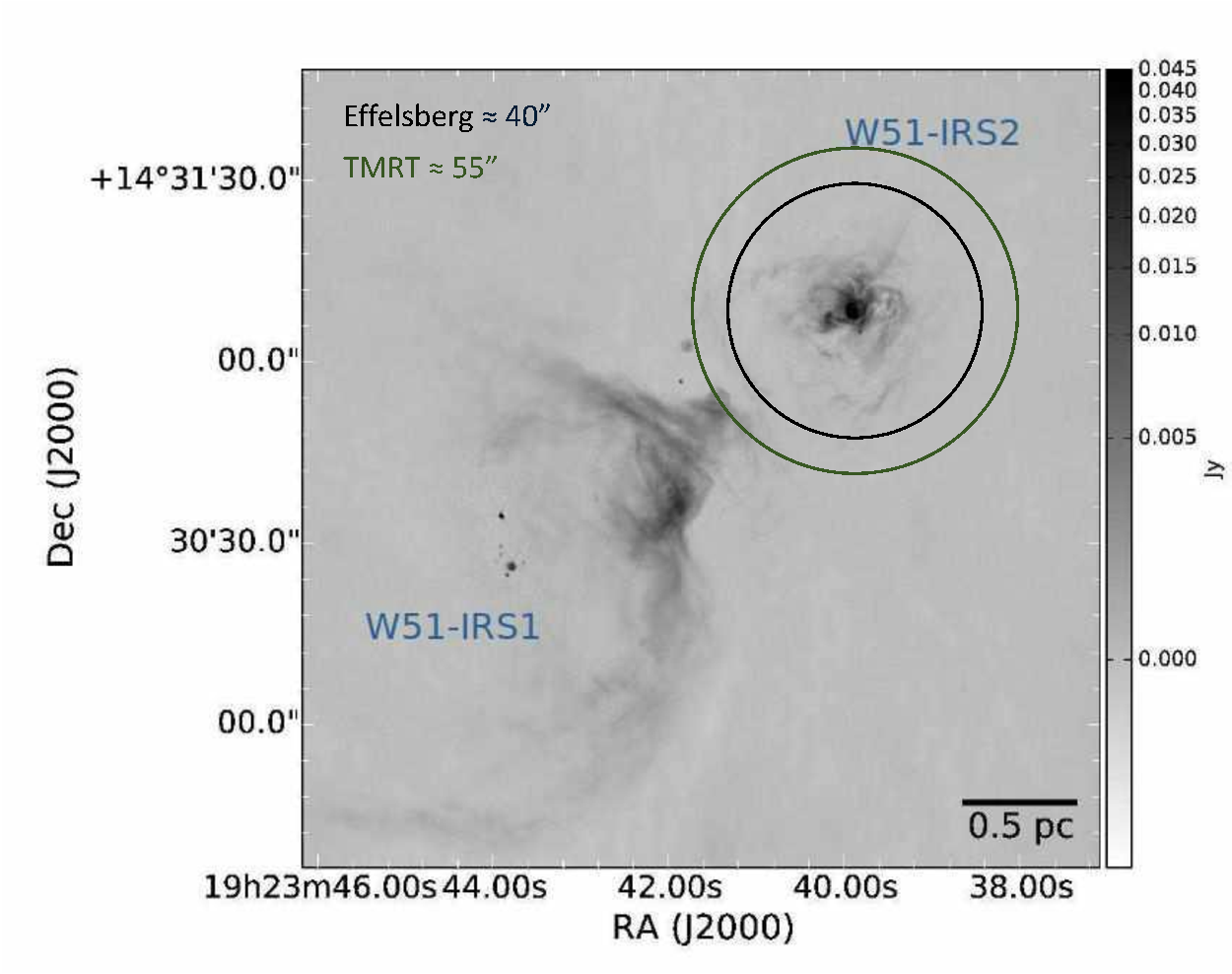}}
\vspace{0.0cm}
\caption{A Jansky Very Large Array continuum map of the W51 region at 14.5\,GHz (image 
taken from Ginsburg et al. 2016; synthesized beam size: 0\ffas33).. W51d is associated with 
the northwestern hotspot, W51-IRS2, while the more extended bright region to the southeast is 
W51-IRS1. Our Effelsberg beam size of $\approx$40$^{\prime\prime}$ (the inner
circle) encompasses the entire IRS2 region, but is small enough to exclude IRS1. The 
slightly larger circle, corresponding to 55$^{\prime\prime}$, refers to the beam size of the 
Tian Ma Radio telescope near Shanghai.}
\label{ginsburg}
\end{figure}

The detection of 11 new maser lines in outer space, all of them from  
NH$_3$ towards the Galactic hot core W51-IRS2 ($D$ $\approx$ 5.0$\pm$0.3\,kpc; 
Sato et al. 2010) and additional masers previously not seen in this specific source 
(Henkel et al. 2013) led to a total of 21 detected ammonia maser transitions alone 
in this hot core, all of them representing inversion lines. For a radio 
continuum image, see Fig.~\ref{ginsburg}, taken from Ginsburg et al. 2016. 
For the detailed distribution of quasi-thermal ammonia emission, see 
Fig.~\ref{goddi}). The single-dish study by Henkel et al. (2013) was complemented 
by observations with much higher angular resolution ($\approx$0\ffas2; Goddi et 
al. 2015) to localize some of the most characteristic maser features and by 
Zhang et al. (2022, 2024), who measured maser lines during five epochs in early 
2020 with the 65-m Tian Ma Radio Telescope (TMRT) and the Jansky Very Large Array (JVLA).

Here we provide follow-up observations of these so far poorly studied maser lines, 
presenting for many of these features spectrally resolved line shapes and information 
on their variability over a time scale of $\ga$10\,yrs. Complementary measurements 
of an ammonia line connecting two vibrationally excited energy levels and of SiO 
$J$ = 1$\rightarrow$0 maser lines are also addressed. Sect.\,2 describes the measurements 
and Sect.\,3 introduces the observed data, while a deeper analysis of measured profiles 
is given in Sect.\,4. Finally, Sect.\,5 summarizes the main results.

\section{Observations}

\subsection{Ammonia (NH$_3$) at K-band ($\lambda$$\approx$1.3\,cm)}

\begin{figure}[t]
\vspace{0.0cm}
\hspace{-0.3cm}
\resizebox{10.4cm}{!}{\includegraphics{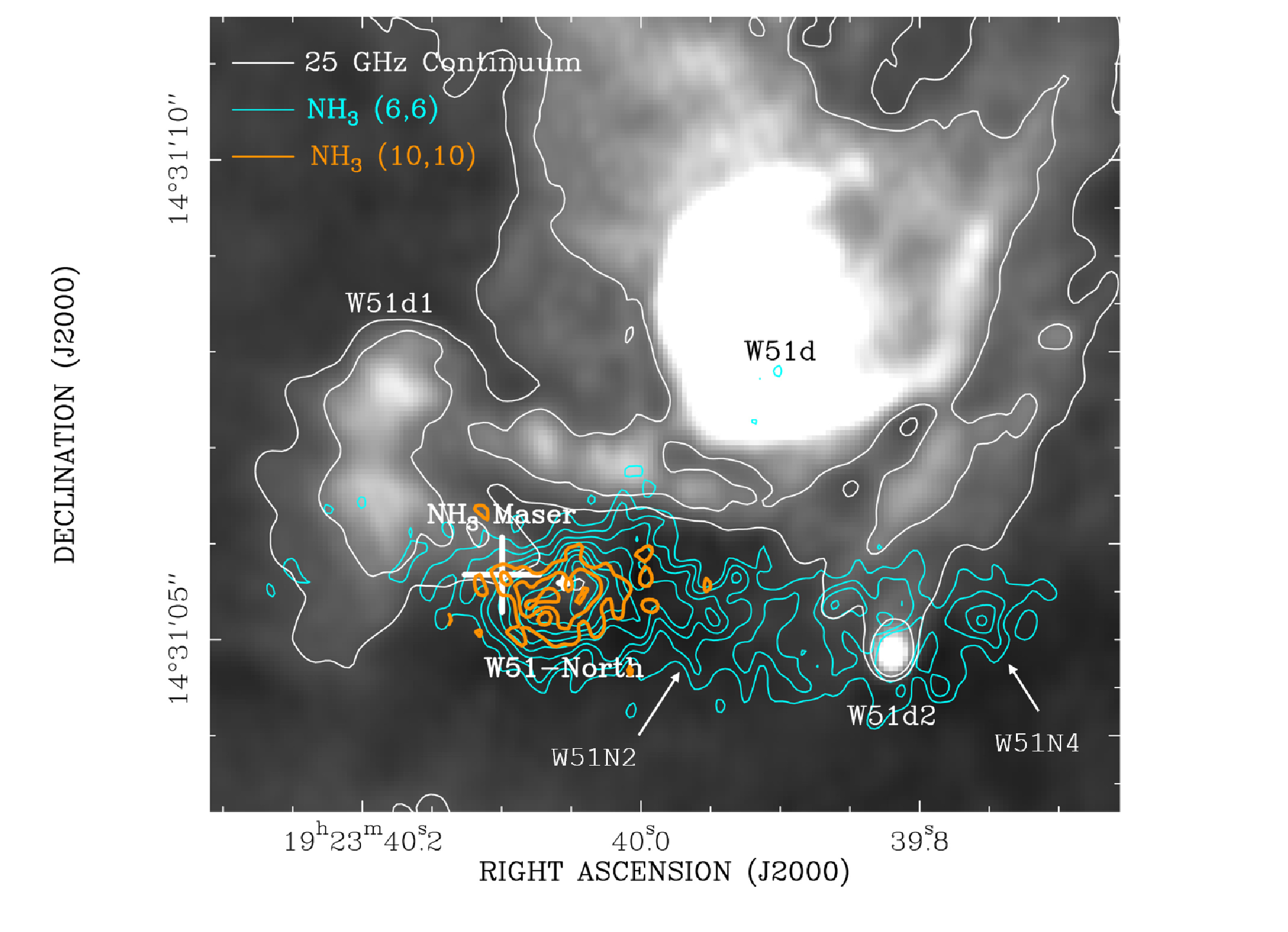}}
\vspace{-0.5cm}
\caption{Overlay of the 25\,GHz continuum emission (gray scale and white contours)
and the integrated intensities of the quasi-thermal (6,6) and (10,10) lines of 
ammonia (taken from Goddi et al. 2015). The (6,6) contours show the elongated 
dense molecular environment at the southern boundary of the H{\sc ii} region 
W51d. The location of main molecular hotspots, W51N4, W51d2, W51N2 and W51-North
(from west to east), are indicated. For more details, see Goddi et al. (2015).} 
\label{goddi}
\end{figure}

New data were taken at several epochs between 2013 January and 2013 May and then 
again in 2017 February, 2018 January and February, 2020 January, and several 
times between 2022 March and 2023 May with the 100-m telescope at 
Effelsberg\footnote{This publication is based on observations with the 
100-m telescope of the MPIfR (Max-Planck-Institut f{\"u}r Radioastronomie) 
at Effelsberg.} near Bonn/Germany. For earlier measurements also playing a role 
in this work, see Henkel et al. (2013). The chosen position was as in Henkel 
et al. (2013) $\alpha_{\rm J2000}$ = 19$^{\rm h}$~23$^{\rm m}$~39\ffs83, 
$\delta_{\rm J2000}$ = 14$^{\circ}$~31$'$~10\ffas1, about 4\ffas7 north of 
W51d2 and ($\Delta\alpha$,$\Delta\delta$) $\approx$ (--2\ffas8,4\ffas7) northwest of 
the main source W51-North (see fig.\,5 of Zhang et al. 2024, our Fig.~\ref{goddi}
and Sect.\,4.6). At the line frequencies observed (18--26\,GHz), the beam size 
is 50$''$ -- 35$''$. Intensity scales were established with a noise diode by 
continuum cross scans towards 3C\,286 and NGC\,7027. Flux densities were taken 
from Ott et al. (1994), also accounting for a 0.5\%\,yr$^{-1}$ secular decrease 
in the case of NGC\,7027, and for gain variations of the telescope as a function 
of elevation \footnote{eff100wiki.mpifr-bonn.mpg.de/doku.php?id=information}. 
1923+210 and 2145+06 were used as pointing sources. The pointing accuracy was 
better than 10$''$. For maser sources much more compact than the beam size, line 
shapes are not affected by pointing errors. 

During the measurements in 2013, the front-end was a dual channel (including the 
two orthogonal linear polarizations) cooled primary focus HEMT (Hot Electron 
Mobility Transistor) receiver with a $T_{\rm sys}$ equivalent to $\approx$65\,Jy
per channel. In 2017 and later, a secondary focus 1.4\,cm two horn system was 
used covering left and right hand circular polarization with a $T_{\rm sys}$ 
equivalent to $\approx$55\,Jy.

For the earlier measurements in 2013, a fast Fourier transform spectrometer (FFTS) 
was employed with a bandwidth of 100\,MHz and 32768 channels per linear polarization,
implying channel widths of order 0.04\,km\,s$^{-1}$. During some of the later measurements 
in 2017 to 2023, the full 8\,GHz wide band of the secondary focus receiver could be covered 
simultaneously by several FFTS spectrometers. For each circular polarization this 
resulted in four partially overlapping spectra with 2.5\,GHz bandwidth and 65536 
spectral channels, centered at 19.25, 21.15, 22.85, and 24.75\,GHz and yielding 
channel widths of order 0.5\,km\,s$^{-1}$. In 2022 and 2023, a high resolution 
backend with 65536 channels and a bandwidth of 300\,MHz was occasionally employed, 
providing a channel width of 0.07\,km\,s$^{-1}$ at 18.5\,GHz. Spectral resolutions are 
slightly coarser than the respective channel widths, by 16\% (see Klein et al. 2012)
for the spectra taken after 2013.

All data for the lines relevant for this article (see Table~1) were taken in a position 
switching mode, employing scans of 5\,min total (on + off) duration and alternating 
offset positions 15$'$ east and west. For the calibration of the data taken in 2013, 
narrow band ($\approx$500\,MHz) continuum scans were taken to account for frequency dependent 
variations of the noise diode signal. For the 8\,GHz wide spectra, ``spectroscopic'' 
pointings covering the entire band were obtained with individual spectral widths of 
$\approx$300\,MHz. The wide ($\approx$2.5\,GHz) spectral subbands were split into smaller 
subbands of size $\approx$300\,MHz, accounting for atmospheric attenuation and elevation 
dependent telescope gain corrections. The calibration uncertainties are estimated to 
be $\pm$15\% after accounting for the elevation dependent gain of the 100-m telescope.

\subsection{Silicon monoxide (SiO) with the 100-m telescope}

In 2014, April, we also searched for $J$ = 1$\rightarrow$0 SiO maser emission in the 
first and second vibrationally excited levels ($v$ = 1 and 2) towards W51-IRS2. As in 
the case of ammonia, position switching with alternating offsets in east-west directions 
was used. These measurements were carried out with a secondary focus receiver with a 
$T_{\rm sys}$ equivalent of $\approx$300\,Jy. For these measurements, an FFTS with 
100\,MHz bandwidth and 32768 channels has been employed. The beam size was 22$''$. 
Pointing was based on measurements of the nearby source 1923+210 and was found to 
be accurate within 5$''$ -- 7$''$. K3-50A was used as calibration source. We adopted 
a flux density of 6.9$\pm$1.0\.Jy for K3-50A (Howard et al. 1997, their fig.~2). 
With the given flux density error and in view of the large difference in declination 
between K3-50A and W51-IRS2, we estimate a calibration uncertainty of $\pm$25\% after 
accounting for elevation dependent telescope gain variations.

\subsection{Ammonia (NH$_3$) with the 12-m APEX telescope} 

We also observed emission from a line of vibrationally excited ammonia towards W51-IRS2, 
selecting the $\nu_2$ = 1, (0,0) $\rightarrow$ (1,0) transition at 466.244\,GHz. 
The measurements were carried out in 2013, April, using APEX\footnote{This publication 
is also based on data acquired with the Atacama Pathfinder Experiment (APEX), which was 
a collaboration between the Max-Planck-Institut f{\"u}r Radioastronomie (MPIfR), the 
European Southern Observatory (ESO), and the Onsala Space Observatory (OSO).} (Atacama 
Pathfinder Experiment; G{\"u}sten et al. 2006), located on the Chajnantor plateau in Chile.  
The front-end was FLASH$^+$ (First Light APEX Submillimeter Heterodyne, Klein et al. 
2014), a dual frequency SiS mixer system operating simultaneously, on orthogonal polarizations, 
in the 345\,GHz and 460\,GHz atmospheric windows. 

The beam size was 13$''$. Observed were the CO $J$ = 3$\rightarrow$2 line at 
$\approx$345.796\,GHz, its image frequency at 333.796\,GHz, as well as ammonia at 
$\approx$466.245\,GHz and its image frequency at 478.245\,GHz. Each of these frequencies 
were covered with two 2.5\,GHz wide spectra, one offset to the blue and the other 
one to the red side, including an overlapping region of width $\approx$0.7\,GHz 
centered at the line frequency. 

The measurements were carried out in cold clear weather with precipitable water vapor measuring
$\approx$0.65\,mm and system temperatures of $T_{\rm sys}$ $\approx$ 200 and 500\,K at 
345\,GHz and 466\,GHz, respectively, on a $T_{\rm A}^*$ scale. Pointing accuracy was established
by measurements of Saturn. The resulting antenna temperatures $T_{\rm A}^*$ were converted to 
main beam brightness temperatures $T_{\rm mb}$ = $T_{\rm A}^*$ $\cdot$\ $F_{\rm eff}$/$B_{\rm eff}$ 
using a forward hemisphere efficiency of $F_{\rm eff}$ = 0.95 and a main beam efficiency of 
$B_{\rm eff}$ = 0.60. The conversion factor between the $T_{\rm A}^*$ temperature and 
the flux density scale is $\approx$50\,Jy/K.

\begin{table} \caption[]{Studied ammonia lines showing at least potential maser features}
\begin{flushleft}
\begin{tabular}{lcrccc}
\hline
      Line     &   $\nu$   &  $E_{\rm up}$ & Species &   V          & Ref. \\ 
     ($J,K$)   &   (MHz)   &      (K)      &         &(km\,s$^{-1}$)&      \\
\hline
      (5,2)    & 20371.450 &     406.7     &   para  &  57          &   1   \\
      (5,3)    & 21285.275 &     380.4     &  ortho  &  57          &   2   \\
      (5,4)    & 22653.022 &     343.3     &   para  &  54+57       &   3   \\
      (6,2)    & 18884.695 &     577.6     &   para  &  54          &   2   \\
      (6,3)    & 19757.538 &     551.3     &  ortho  &  57          &   4   \\
      (6,4)    & 20994.617 &     514.3     &   para  &  57          &   5   \\
      (6,6)    & 25056.025 &     408.1     &  ortho  &  45          &   6   \\
      (7,3)    & 18017.337 &     750.2     &  ortho  &  54          &   5   \\
      (7,4)    & 19218.465 &     713.5     &   para  &  54          &   2   \\
      (7,5)    & 20804.830 &     666.0     &   para  &  54          &   3   \\
      (7,6)    & 22924.940 &     607.8     &  ortho  &  45+50       &   2   \\
      (7,7)    & 25715.182 &     538.6     &   para  &  45+50       &   2   \\
      (8,5)    & 18808.507 &     893.3     &   para  &  54          &   2   \\
      (8,6)    & 20719.221 &     835.5     &  ortho  &  45+54       &   7   \\
      (9,6)    & 18499.390 &    1090.9     &  ortho  &  45 -- 67    &   4   \\
      (9,7)    & 20735.452 &    1022.6     &   para  &  54          &   2   \\
      (9,8)    & 23657.471 &     943.3     &   para  &  54          &   3   \\  
     (10,7)    & 18285.434 &    1306.1     &   para  &  54          &   2   \\
     (10,9)    & 24205.287 &    1137.6     &  ortho  &  45+50       &   2   \\
     (11,9)    & 21070.739 &    1449.9     &  ortho  &  45+50+54    &   7   \\
\hline
\end{tabular}
\end{flushleft}
{\it Notes}: Frequencies (column 2) and upper state energies $E_{\rm up}$
(column 3) are taken from the Jet Propulsion Laboratory (JPL) molecular 
line catalog (Pickett et al. 1998) and the Cologne Database for Molecular 
Spectroscopy (CDMS; M{\"u}ller et al. 2005; Endres et al. 2016). Differences
between lower and upper state energies correspond to approximately 1\,K.
Column 4 provides the ammonia species (ortho or para), while column 5
shows the approximate velocities of the maser components. Note that 
the term 45\,km\,s$^{-1}$ has been taken from Henkel et al. (2013). Because 
of a positive velocity drift, actual velocities tend to be higher by 
typically 2--4\,km\,s$^{-1}$. The last column provides a reference to the 
first detection of the respective maser line in the interstellar medium.
1: This paper; 2: Henkel et al. (2013); 3: Mauersberger et al. (1987);
4: Madden et al. (1986); 5: Mei et al. (2020); 6: Beuther et al. (2007); 
7: Walsh et al. (2007)a.
\label{abc}
\end{table}

\section{Results}

\subsection{General aspects}

Making use of the available 8\,GHz wide Effelsberg spectra, we checked systematically 
all 49 ammonia inversion transitions with rotational quantum numbers up to $J$ = 16
that can be found in the observed frequency interval between 18 and 26\,GHz. 
Figs.~\href{zenodo.org/records/15746097}{A.1} to \href{zenodo.org/records/15746097}{A.44}
in Appendix A show the detected NH$_3$ inversion lines in the ground vibrational level 
with at least tentatively detected maser features. 20 lines following rising quantum 
number $J$ from $J$ = 5 to 11 are included, with a spectrum from 2012 (also shown by 
Henkel et al. 2013) at the bottom left and the most recent ones in the upper right. 
For the particularly frequently measured (9,6) and (10,7) lines, new spectra are 
presented exclusively. Characteristic channel widths are 0.5 -- 1.0\,km\,s$^{-1}$. 
While a comparison of spectra from 2012 (also published by Henkel et al. 2013), 2013, 
2017, 2018, 2020, 2022 and 2023 provides a view on long term variability, we also 
present (if lines are strong enough) data at five epochs within a six day interval 
between 2018 January 24 and 2018 February 1 to trace variability on short time scales. 
Covering slightly larger time intervals, we also show spectra from five monitoring 
epochs between 2023 April and May 4. Furthermore, a large number of high spectral 
resolution ($J,K$) = (9,6) and (10,7) profiles are presented, taken in 2022 and 
2023, while a few high resolution spectra from other inversion lines are also obtained. 
Results from Gaussian fits are given in Tables~B.1 to B.19 of Appendix B.

\subsection{Use of the quasi-thermal lines}

While this article focuses on maser features, the quasi-thermal emission centered near 
$V_{\rm LSR}$ = 60\,km\,s$^{-1}$ (here and elsewhere we use Local Standard of Rest 
velocities) is useful because it arises from relatively large volumes and may 
thus not show significant variability. As a consequence the thermal features can be used 
as an additional means to calibrate spectra, providing an independent check with respect 
to the calibration methods outlined in Sect.\,2.1. Comparing our quasi-thermal features 
with those reported by Mauersberger et al. (1987; their Table~1, 1.4\,K in units of 
main beam brightness  temperature correspond to about 1\,Jy), which were also obtained 
with the Effelsberg 100-m telescope, agreement is reasonably good in case of the (5,4), 
(7,6), (7,7), (9,7), and (10,8) lines. Furthermore, our detected 60\,km\,s$^{-1}$ feature 
of the (6,2) line is consistent with the upper limit provided by Mauersberger et al. 
(1987). For the (6,6) line, our 60\,km\,s$^{-1}$ feature is consistent with that 
published by Henkel et al. (2013). In the case of the (7,5) line, the quasi-thermal 
feature reaches levels of 0.05 -- 0.08\,Jy, slightly below the 0.1\,Jy reported 
by Mauersberger et al. (1987). A similar situation is encountered in the case of 
the (8,6) line, while our calibration yields quasi-thermal (9,8) and (11,9) features about 
a factor of two weaker than those reported by Mauersberger et al. (1987). To summarize, our 
calibrated spectra provide flux density scales that have a tendency to yield smaller values 
than those reported by Mauersberger et al. (1987), ranging from good agreement in most cases
down to a factor of two below the previously reported peak intensities in the most extreme cases.
Discrepancies tend to appear in the lines with highest excitation above the ground state, i.e. 
in those cases where the quasi-thermal emission centered near $V_{\rm LSR}$ = 60\,km\,s$^{-1}$
is showing a particularly compact spatial distribution (Goddi et al. 2015). Interferometric 
data (Fig.~\ref{goddi} and again Goddi et al. 2015) indicate sizes of 
$\approx$0.5$\times$0.1 and $\approx$0.1$\times$0.1\,lightyears$^2$ for the (6,6) and (9,9) 
lines and suggest even smaller sizes for the non-metastable inversion lines, so that 
variability within a decade may well be possible. Nevertheless it remains open whether 
the observed discrepancies in peak intensity are caused by pointing errors, calibration 
uncertainties or by real changes in the emission profile. To stay on the conservative 
side, in the following it is assumed that all discrepancies in the time interval between 
2012 and 2023 are due to calibration uncertainties.

\subsection{New maser lines} \label{new_masers}

\begin{figure}[t]
\vspace{-0.0cm}
\resizebox{24.0cm}{!}{\rotatebox[origin=br]{-90}{\includegraphics{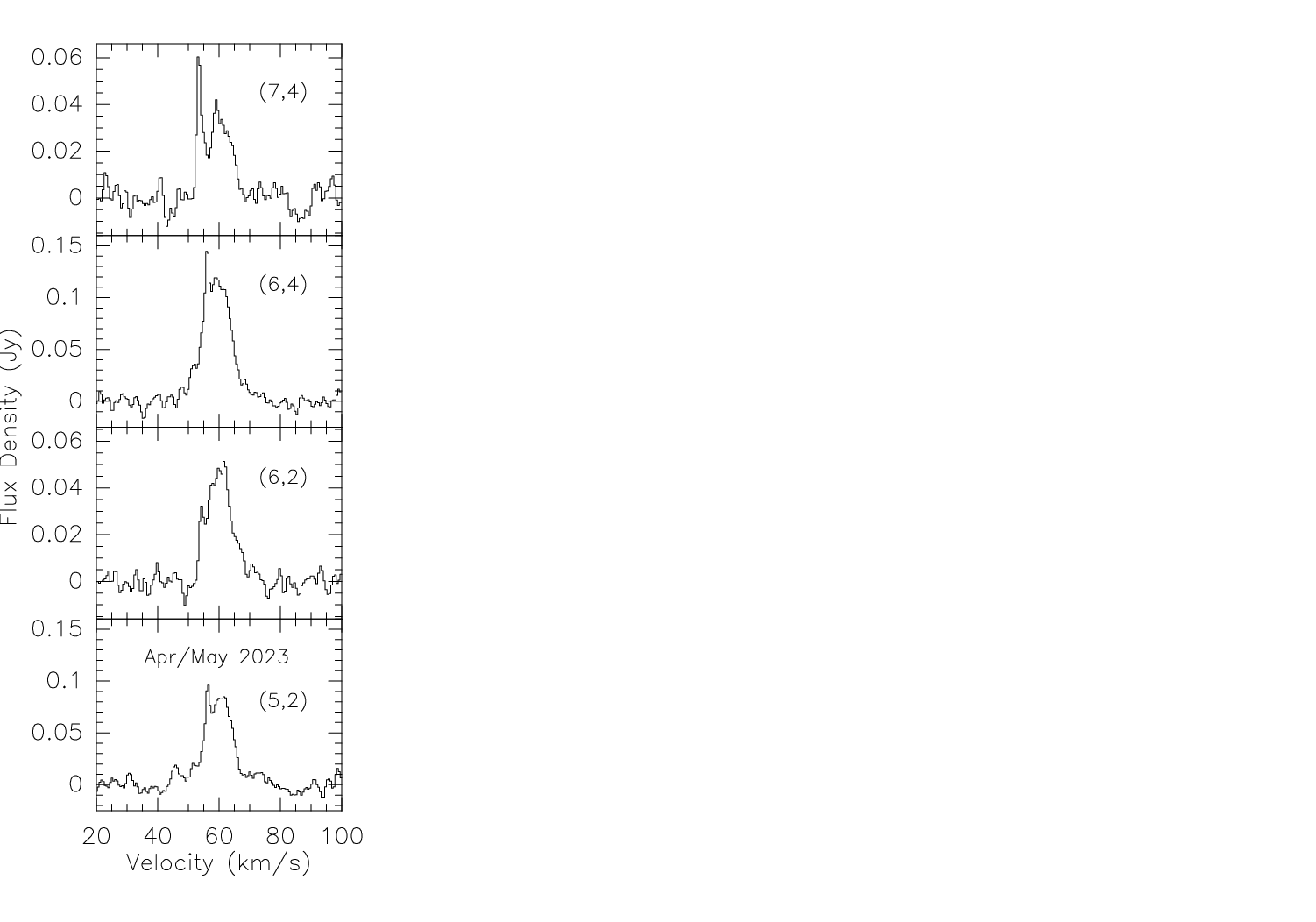}}}
\vspace{-1.3cm}
\caption{Averaged NH$_3$ spectra from April/May 2023. Channel widths are 0.60, 0.55, 
0.61, and 0.56\,km\,s$^{-1}$ from top to bottom, respectively.} 
\label{2023a}
\end{figure}

Searching for new maser transitions not yet seen in W51-IRS2 or, even better, not yet seen 
anywhere in the interstellar medium, we have analyzed the 49 inversion lines mentioned
in Sect.~3.1 and have found two candidates. The first is the (5,2) transition 
(Fig.~\href{zenodo.org/records/15746097}{A.1}). While apparently not being present in 2012, 
2017 and 2020, there is a feature off the quasi-thermal $V_{\rm LSR}$ = 60\,km\,s$^{-1}$ 
component in late April and early May 2023, not being present in early April and late May 2023. 
(5,2) maser emission has to our knowledge never been reported before (see, e.g., table A.1 
in Yan et al. 2024). Another maser candidate is the (6,4) inversion line 
(Figs.~\href{zenodo.org/records/15746097}{A.11} and \href{zenodo.org/records/15746097}{A.12}), 
which is seen throughout April and early May 2023, disappearing like the (5,2) feature in 
late May. Velocities are in both cases $V_{\rm LSR}$ $\approx$ 57\,km\,s$^{-1}$. Note that 
a (6,4) maser was reported before by Mei et al. (2020) in Sgr\,B2(N). Signal-to-noise ratios f
or both spectral features, in the (5,2) and (6,4) lines, are such that we rate them as 
tentatively detected following Figs.\,\href{zenodo.org/recorde/15746097}{A.1}, 
\href{zenodo.org/records/15746097}{A.11} and \href{zenodo.org/records/15746097}{A.12}. However, the 
averages of the spectra taken in April and May 2023 (Fig.\,\ref{2023a}) show those 
narrow features more convincingly with signal-to-noise ratios of 3.5 and 10 (Tables~B.7 
and B.10) that are most readily interpreted in terms of maser emission. While 10$\sigma$
may appear to be high for a tentative detection, we should keep in mind that the feature is heavily 
blended by the comparatively broad quasi-thermal feature centered at $\approx$60\,km\,s$^{-1}$. 
The (7,3) maser, not covered by Henkel et al. (2013) and being first detected by Zhang et al. (2022), 
is confirmed and demonstrated to be present since at least 2018. This implies that during the last 
decade no strong new maser line has appeared with the potential exception of the (7,3) transition.

\begin{table} \caption[]{NH$_3$ maser families in W51-IRS2}
\begin{flushleft}
\begin{tabular}{ccc}
\hline
     Velocity    &  Transitions                                              & Notes \\ 
   (km\,s$^{-1}$ &   ($J,K$)                                                 &       \\
\hline
    45           & (6,6),(7,6),(7,7),(8,6),(10,9),(11,9)                     & 1     \\
    54           & (6,2),(7,3),(7,4),(7,5),(8,5),(9,7),(9,8),(10,7)          & 2     \\
    57           & (5,2),(5,3),(5,4),(6,3),(6,4)                             &       \\
  Other          & (9,6)                                                     & 3     \\
\hline
\end{tabular}
\end{flushleft}
{\it Notes}: 1 -- Due to a secular drift towards higher velocities, now the formerly
termed 45\,km\,s$^{-1}$ component is seen at slightly higher velocities. 2 -- The 
(5,4) line also contains a weak feature near 54\,km\,s$^{_1}$, but maser emission 
at 57\,km\,s$^{-1}$ dominates. Some other transitions also show occasionally double
maser features. These are addressed in Sect.\,3.4.4. 3. -- The (9,6) line covers a 
wide velocity range from $<$50 to $>$60\,km\,s$^{-1}$, see 
Figs.\,\href{zenodo.org/records/15746097}{A.29} to \href{zenodo.org/records/15746097}{A.32}.  
\label{family}
\end{table}

\subsection{The maser families}

Henkel et al. (2013) distinguished between four maser families, those measured near 
45\,km\,s$^{-1}$ (or at slightly higher velocities), 54\,km\,s$^{-1}$, and 
57\,km\,s$^{-1}$ and the (9,6) maser line, which is unique with respect to its 
strength and its wide velocity range. We trace five transitions definitely showing 
the 57\,km\,s$^{-1}$ maser component (the (5,2), (5,3), (5,4), (6,3) and (6,4) lines), 
while six host features near 45\,km\,s$^{-1}$ (the (6,6), (7,6), (7,7), (8,6), (10,9), 
and (11,9) transitions). In these cases occasionally even two velocity components are 
found of which the lower velocity feature is addressed in Sect.\,3.4.1. The phenomenon 
is discussed in more detail in Sect.\,3.4.2. Eight transitions emit maser emission 
near 54\,km\,s$^{-1}$, the (6,2), (7,3), (7,4), (7,5), (8,5), (9,7), (9,8), and (10,7) 
lines. In the (5,4) line a weak feature accompanying the main 57\,km\,s$^{-1}$ maser 
component is also seen near 54\,km\,s$^{-1}$ (Fig.~\href{zenodo.org/records/15746097}{A.6}). 
The maser families are summarized in Table~\ref{family}.

\begin{figure}[t]
\vspace{0.0cm}
\resizebox{24.0cm}{!}{\rotatebox[origin=br]{-90}{\includegraphics{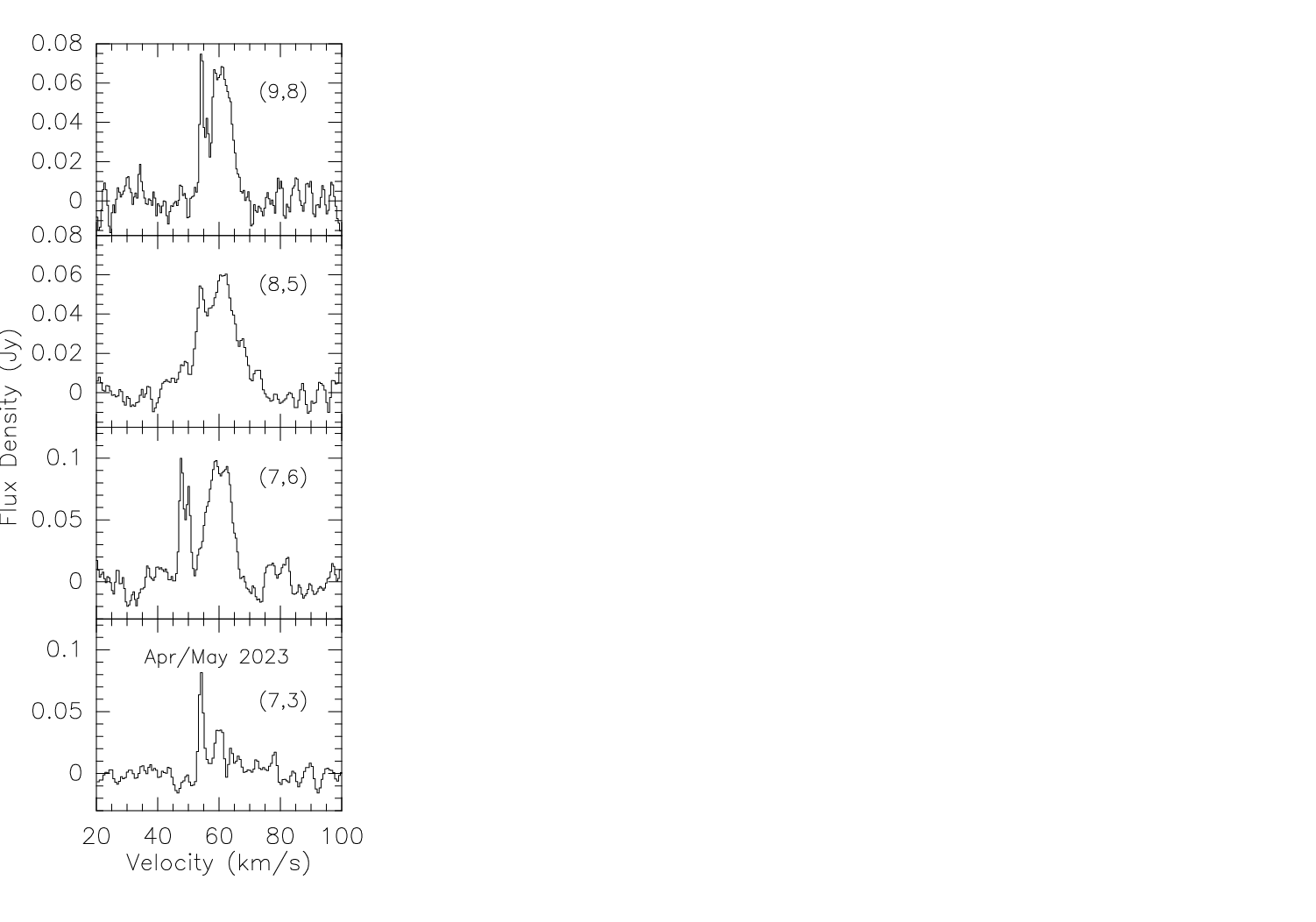}}}
\vspace{-1.3cm}
\caption{Averaged NH$_3$ spectra from April/May 2023. Channel widths are 0.48, 0.61,
0.50. and 0.64\,km\,s$^{-1}$ from top to bottom.} 
\label{2023b}
\end{figure}

\begin{figure}[t]
\vspace{0.0cm}
\resizebox{24.0cm}{!}{\rotatebox[origin=br]{-90}{\includegraphics{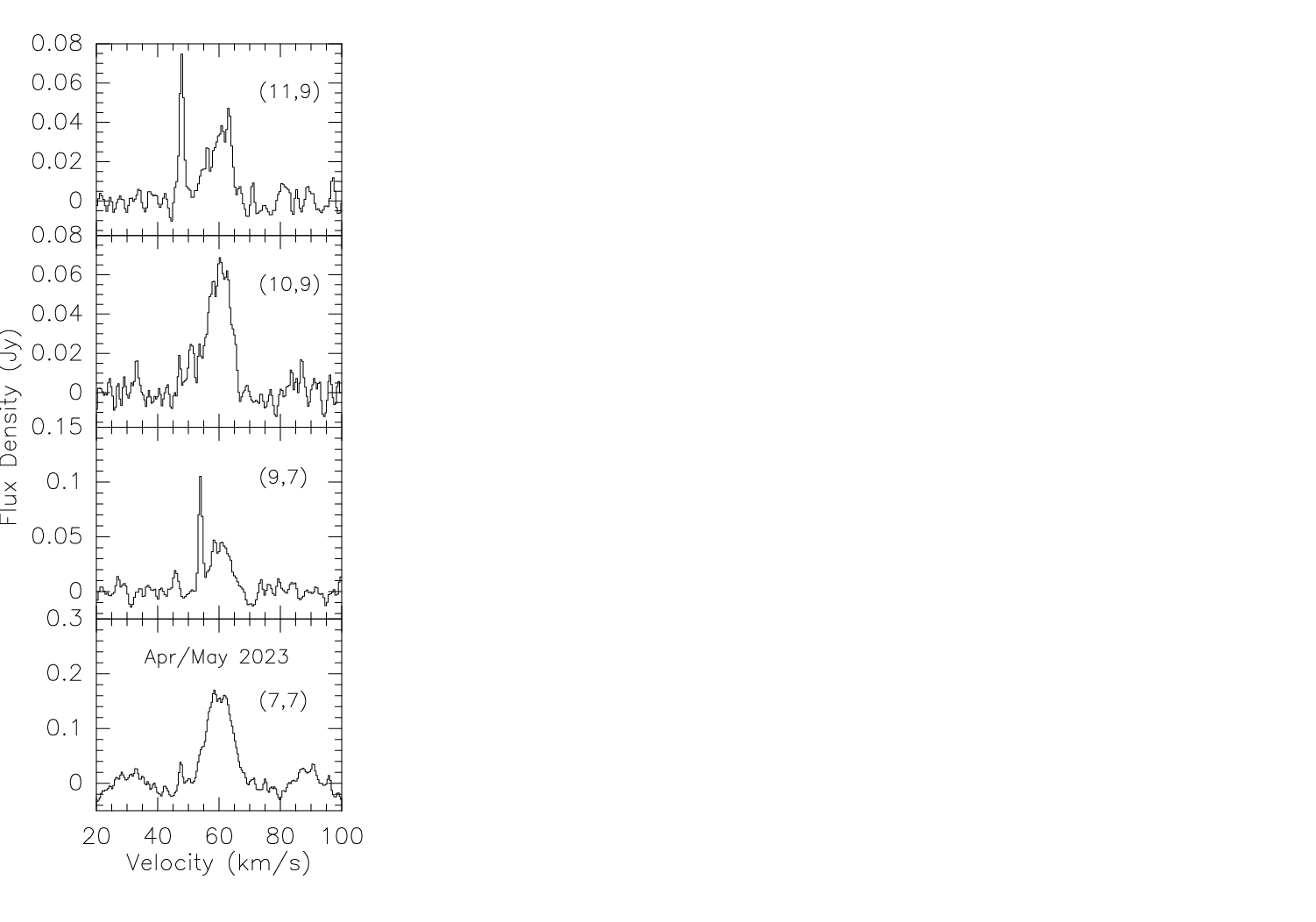}}}
\vspace{-1.3cm}
\caption{Averaged NH$_3$ spectra from April/May 2023. Channel widths are 0.54, 0.47,
0.55, and 0.45\,km\,s$^{-1}$ from top to bottom, respectively.}
\label{2023c}
\end{figure}

\subsubsection{Variability on comparatively long timescales}

The first question is related to variability. Are these correlated, in the 
sense that all the masers in Figs.~\href{zenodo.org/records/15746097}{A.1} 
to \href{zenodo.org/records/15746097}{A.44} change their flux densities in a 
similar manner or are at least those belonging to a specific maser family correlated? Unlike 
in the case of the comparatively broad quasi-thermal profiles, maser lines tend to be 
narrow so that, to be independent from the chosen channel spacing, integrated intensities 
will be compared. For the moment neglecting the (9,6) line and starting with the lowest 
velocity component, that near 45\,km\,s$^{-1}$, we see an almost general decline in all six 
traced transitions. Flux densities from April 2012 to early 2018 decrease by factors of 
order 40, 5, 1.0, 10, 10 and 50 for the (6,6), (7,6), (7,7), (8,6), (10,9), and (11,9)
transitions, respectively. Averaging our (6,6) profiles from 2023 April and May does 
not yield a detection with a 3$\sigma$ value of 0.1\,Jy for a channel spacing of 
0.46\,km\,s$^{-1}$. For the eight lines near 54\,km\,s$^{-1}$ we obtain a decline
by a factor of 4 for the (6,2) line, little change in the (7,3), (7,4), (7,5) and (8,5)
transitions, a drop by a factor of approximately 1.5 in the (9,7) and (9,8) lines
and a strong increase in the (10,7) line, by a factor of $\approx$15. In case of our three
transitions already previously known to belong to the 57\,km\,s$^{-1}$ maser family, 
changes are not large. Flux densities tend to stay constant in case of the  
(5,3), (5,4) and (6,3) lines between 2012 and 2018.  The (5,2) and (6,4) lines are 
too weak to check variability.

For the time interval between 2018 and 2023, emission from the V$_{\rm LSR}$ $\approx$ 
45\,km\,s$^{-1}$ component is weak. While the (7,7) feature remains strong till 
2020, after 2020 this maser component is only detected in the (7,6), (8,6) and 
(11,9) transitions. The 54\,km\,s$^{-1}$ feature shows quite different trends. 
The (7,3), (7,4), (7,5), (9,7) and (10,7) transitions keep approximately their 
intensities. The (6,2) feature is barely seen in 2023 (Fig.\,\ref{2023a}) and the (8,5) 
line intensity drops by a factor of approximately three between 2018 and 2020 and remains 
weak (and constant) till the end of our monitored time interval (Fig.\,\ref{2023b}). The 
(9,8) transition is weak but still present throughout this time period (Fig.\,\ref{2023b}). 

While more details will be given below, the (9,6) line spectra with 
coarse velocity resolution (Figs.\,\href{zenodo.org/records/15746097}{A.29} and 
\href{zenodo.org/records/15746097}{A.30}) show a drift of the strongest velocity 
component towards lower velocities from almost 55 to a little above 53\,km\,s$^{-1}$ 
between 2013 and 2023. Its $\approx$60\,km\,s$^{-1}$ component shows a line shape very 
similar to the quasi-thermal features seen in other inversion lines. However, this 
$V_{\rm LSR}$ $\approx$ 60\,km\,s$^{-1}$ component is $\approx$50 times stronger than the 
corresponding features in the (9,7) and (9,8) lines and thus far too strong to be of 
quasi-thermal origin. Near the end of our monitoring time interval, emission is also seen
at velocities $<$50\,km\,s$^{-1}$.

\subsubsection{The most likely pumping scheme}

From this admittedly superficial analysis of variability we can already conclude that 
the different maser families are not directly connected. Even within the families, 
strong differences in variability may occur, but only in one case, that of the 
54\,km\,s$^{-1}$ component, we find both masers with clearly decreasing and 
increasing flux densities. Henkel et al. (2013) and  Goddi et al. (2015) interpreted 
maser emission detected at the same radial velocity, typically encountered in ($J$,$K$) 
and in ($J$+1,$K$) transitions, as caused by infrared pumping. Because of similar 
parities (see, e.g., fig.~10 of Henkel et al. 2013) in adjacent inversion doublets
within a given K-ladder, spontaneous decay through rotational transitions from a ($J$+1,$K$) 
to a ($J$,$K$) inversion doublet leads from the upper state in the ($J$+1,$K$) to the 
lower state in the ($J$,$K$) doublet and vice versa. Thus population inversion in 
the upper inversion doublet would lead to anomalous cooling of the level populations 
in the lower doublet, leading for example with inversion in the (9,7) to anti-inversion 
in the (8,7) and once again to inversion in the (7,7) doublets. To obtain population 
inversion and maser emission in adjacent ($\Delta J$ = 1, $\Delta K$ = 0) doublets, 
vibrational excitation is required through $\approx$10\,$\mu$m photons, abundant in 
the $\approx$300\,K environment of W51-IRS2 (Mauersberger et al. 1987). Then parities 
are reversed twice, from plus to minus and back to plus or from minus to plus and 
then back to minus, first due to radiative excitation by infrared photons and second 
due to rapid spontaneous decay. In this case, inversion in the upper vibrational ground 
state doublet also supports inversion in the lower one.

\begin{figure}[t]
\vspace{0.0cm}
\resizebox{8.0cm}{!}{\rotatebox[origin=br]{0}{\includegraphics{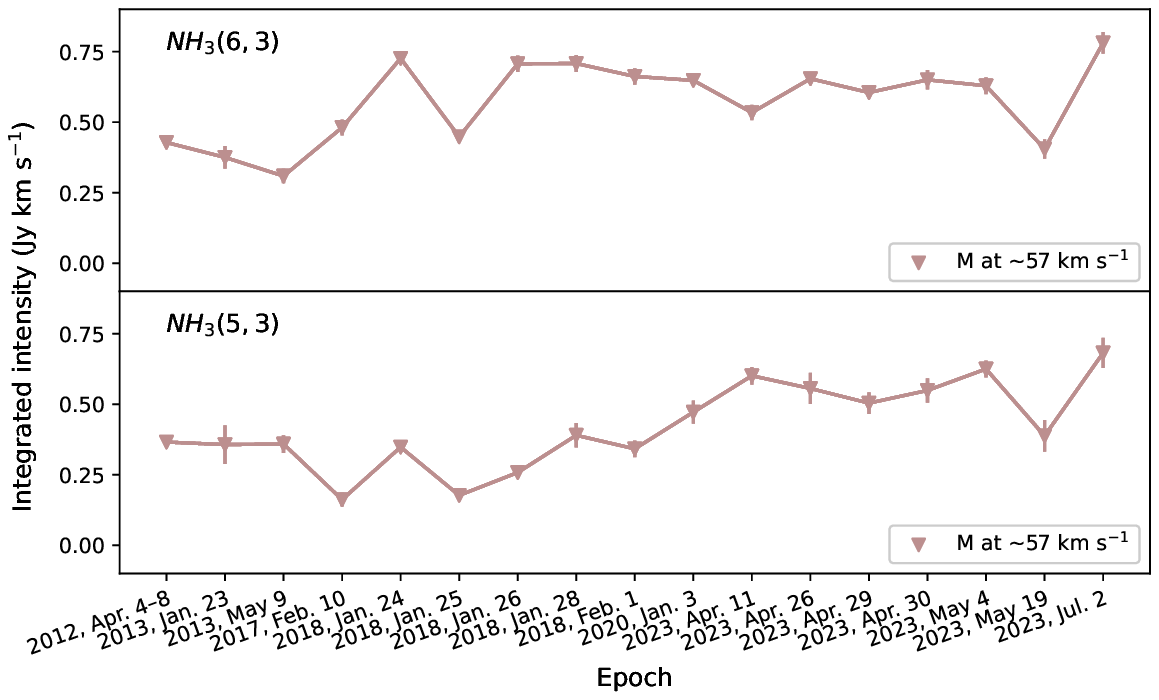}}}
\vspace{0.0cm}
\caption{Comparison of integrated intensities of the (5,3) and (6,3) maser features.
``M'' denotes ``maser'', for which the velocity is given. Note that here and in the 
following plots the timescale (x-axis) is not linear but is designed with equal spacings 
between consecutive observing epochs. Errors are standard deviations from Gaussian
fits.}
\label{53-63}
\end{figure}

\begin{figure}[t]
\vspace{0.0cm}
\resizebox{8.0cm}{!}{\rotatebox[origin=br]{0}{\includegraphics{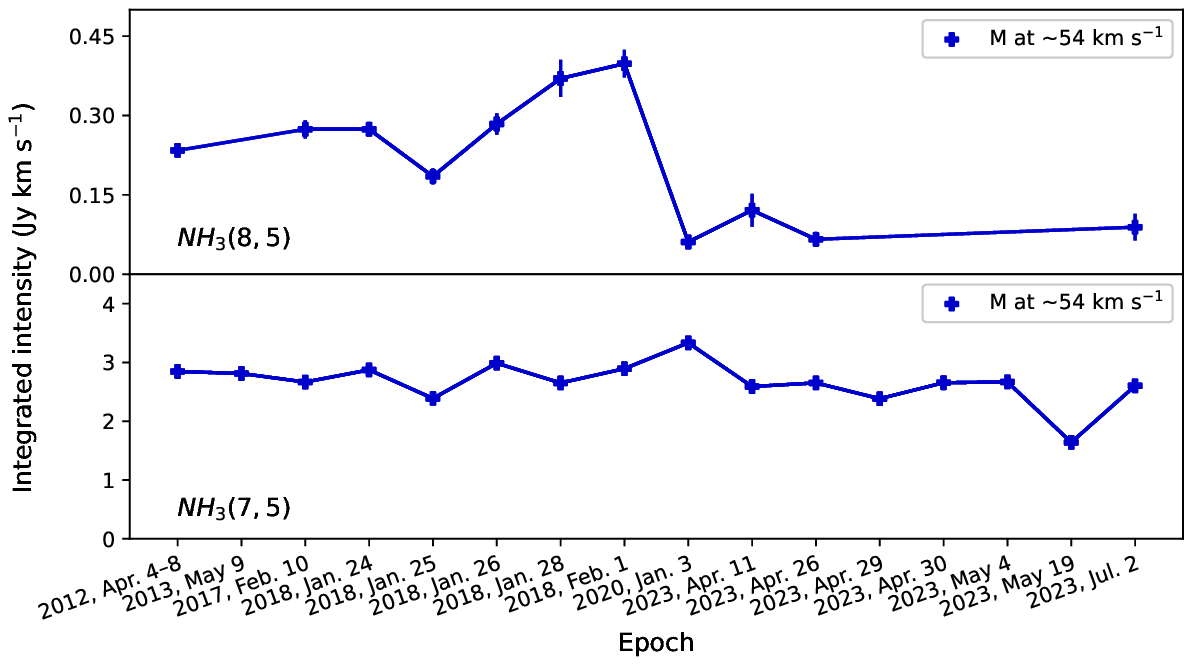}}}
\vspace{0.0cm}
\caption{Comparison of integrated intensities of the (7,5) and (8,5) maser features.
For more details, see Fig.~\ref{53-63}.}
\label{75-85}
\end{figure}

\begin{figure}[t]
\vspace{0.0cm}
\resizebox{8.0cm}{!}{\rotatebox[origin=br]{0}{\includegraphics{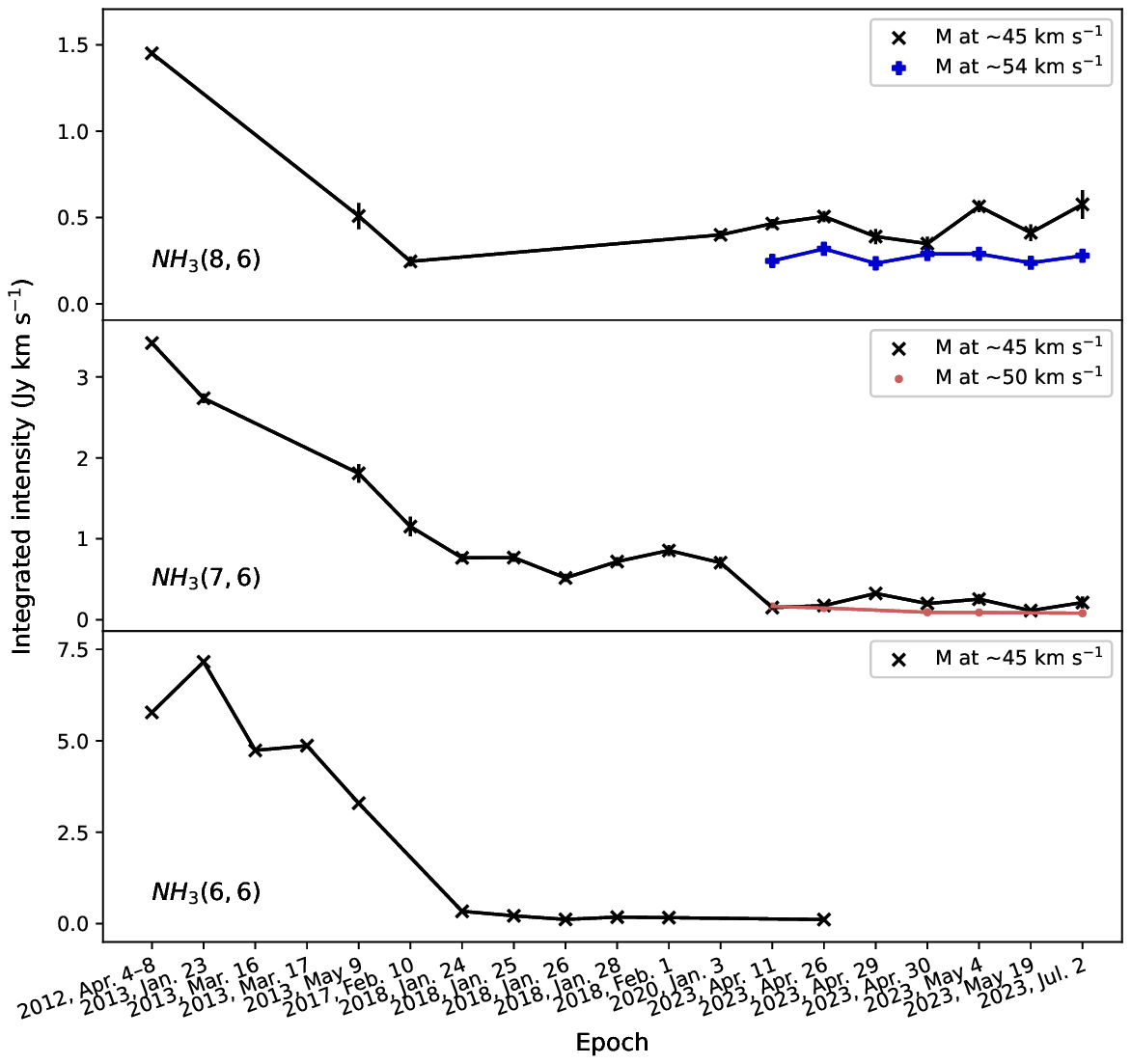}}}
\vspace{0.0cm}
\caption{Comparison of integrated intensities of the (6,6), (7,6) and (8,6) maser features.
For more details, see Fig.~\ref{53-63}.}
\label{66-76-86}
\end{figure}

\begin{figure}[t]
\vspace{0.0cm}
\resizebox{8.0cm}{!}{\rotatebox[origin=br]{0}{\includegraphics{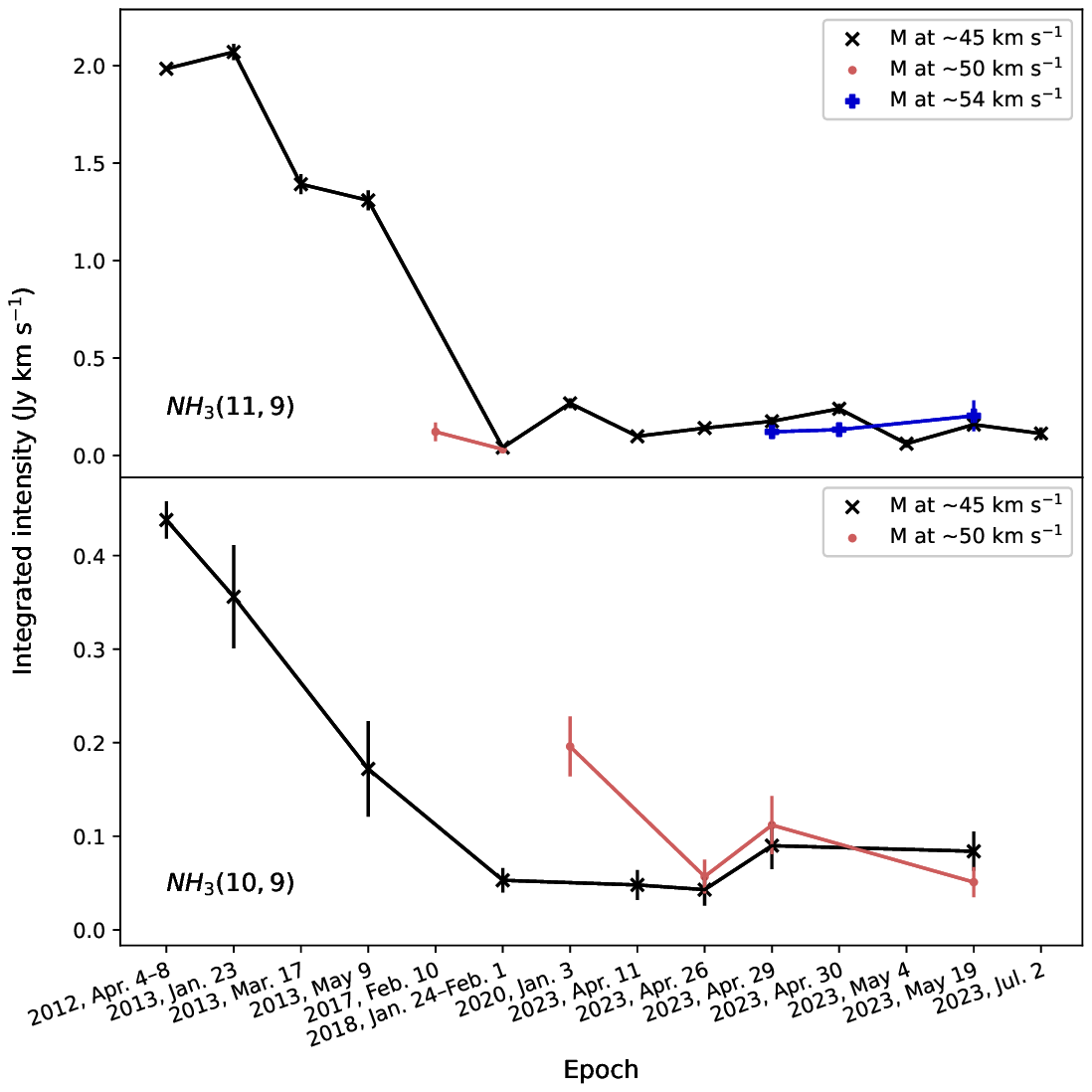}}}
\vspace{0.0cm}
\caption{Comparison of integrated intensities of the (10,9) and (11,9) maser features.
For more details, see Fig.~\ref{53-63}.}
\label{109-119}
\end{figure}

\begin{figure}[t]
\vspace{0.0cm}
\resizebox{8.0cm}{!}{\rotatebox[origin=br]{0}{\includegraphics{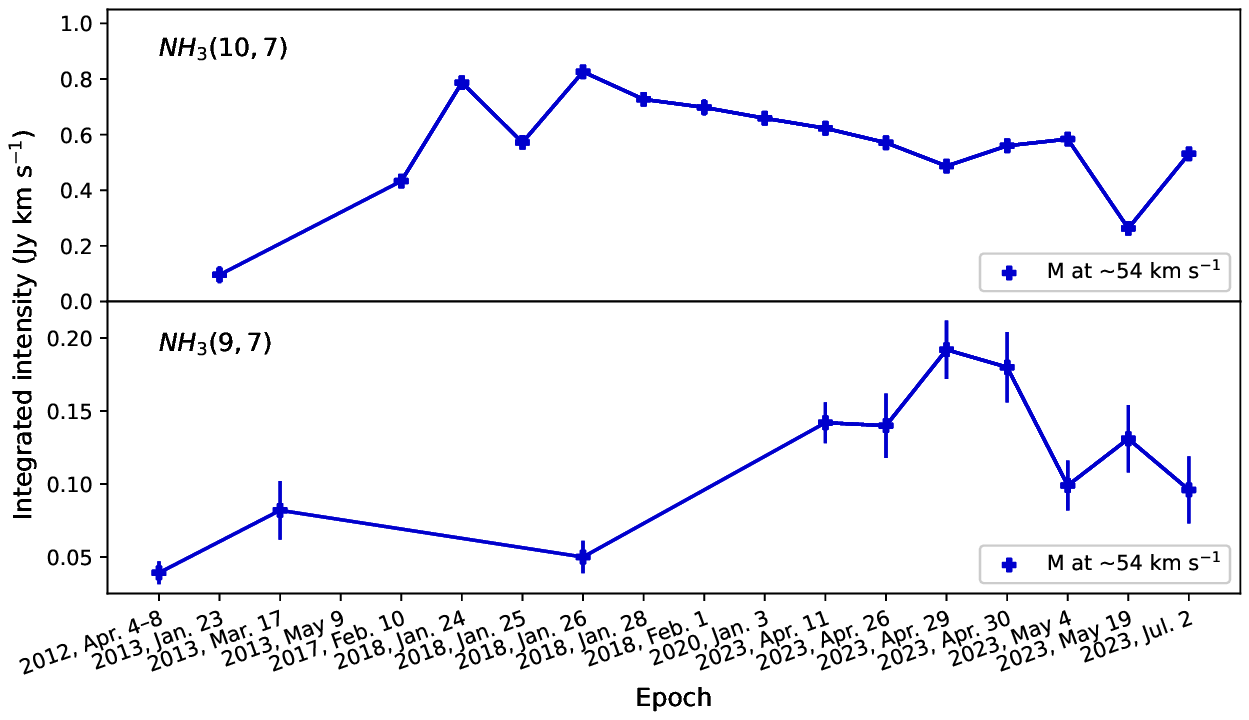}}}
\vspace{0.0cm}
\caption{Comparison of integrated intensities of the (9,7) and (10,7) maser features.
For the (9,7) line, the average for the epochs between 2018 January 24 and February 1
was set to January 26. For more details, see Fig.~\ref{53-63}.}
\label{97-107}
\end{figure}

Our maser lines include a number of adjacent ($J$,$K$) -- ($J$+1,$K$) transitions in a 
given K-ladder, the (5,2) and (6,2), the (9,7) and (10,7), and the (10,9) and (11,9) 
doublets, the (5,3), (6,3) and (7,3) as well as the (5,4), (6,4) and (7,4) triplets, 
and the (6,6), (7,6), (8,6) and (9,6) quadruplet. Thus more than a dozen of our 20 monitored 
maser lines are involved. Furthermore we may look for a connection between the (7,7) and 
(9,7) lines even though the (8,7) line did not reveal any masing component at a $\ga$40\,mJy 
level and the (7,7) line arises from a metastable doublet, being the lowest one in the 
$K$ = 7 stack. This is because in case pumping into vibrational levels does not take place,
a maser line in a ($J$+2,$K$) inversion doublet may lead to anti-inversion in the ($J$+1,$K$)
transition, but may cause once more inversion in the ($J$,$K$) doublet.

First of all we have to check whether the masers in the supposedly connected transitions 
belong to the same family (see Sect.\,3.4). This is indeed the case for all the 
(5,3)--(6,3), (5,4)--(6,4), (7,5)--(8,5), (9,7)--(10,7) and (10,9)--(11,9) maser pairs 
and also for the (6,6), (7,6), and (8,6) maser triple. However, it does not hold for the 
(5,2)--(6,2), (6,3)--(7,3) and (6,4)--(7,4) pairs, which contain features belonging to 
different maser families.  The (9,6) line also shows, like the (6,6), (7,6) and (8,6) 
triple, maser emission near 45\,km\,s$^{-1}$. The only possibility for a maser doublet 
not involving vibrational excitation is the (7,7) and (9,7) pair of lines, with the 
(8,7) transition not showing notable inversion. However, the narrow (7,7) and (9,7) 
maser features are encountered at different velocities and are thus not related. To
summarize: Since the (7,7) and (9,7) masers are not related, all evidence points to 
infrared vibrational pumping and its signature is directly seen in 70\% of our 20 maser 
lines.
 
Having checked, which inversion line pairs provide consistent kinematics, the following 
more detailed analysis also reveals that variability is often but not always correlated. 
For the (5,3) and (6,3) lines no great changes in intensity are seen (Fig.\,\ref{53-63}). 
The (5,4) and (6,4) transitions do not allow for any conclusion, because the tentatively
detected (6,4) maser feature is too weak (Figs.\,\href{zenodo.org/records/15746097}{A.11} 
and \href{zenodo-org/15746097}{A.12}). While the (7,5) maser remains approximately 
constant, the corresponding velocity component in the higher excited (8,5) line drops 
between 2018 and 2020 by a factor of three, but still remains detectable after 
weakening by approximately an order of magnitude between 2012 and 2023 (Figs.\,\ref{2023b} 
and \ref{75-85}). Considering the maser triple, the (6,6) and (7,6) lines become much 
weaker after May 2013, while the (8,6) peak intensity drops already in May 2013 
(Figs.~\ref{66-76-86} and \href{zenodo.org/records/15746097}{A.27}). A (6,6) line 
maser is not detected in 2023, while the two non-metastable maser lines, the (7,6) 
and (8,6) transitions, are still observed. The (10,9) line is getting substantially 
weaker in very early 2013, while this occurs only after May 2013 in case of the higher 
excited (11,9) transition (Fig.~\ref{109-119}). The (9,7) and (10,7) transitions behave 
in a different way: the former (rather weak) maser appears to be strongest in 2023, 
while the latter becomes much stronger during the start of the monitoring time span 
to get slightly weaker more recently (Fig.\,\ref{97-107}). To summarize, while we 
find clear connections with respect to velocity, supporting vibrational excitation, 
variability in most (albeit not all) maser pairs shows individual characteristics.

\subsubsection{Variability on shorter timescales}

As already noted in Sect.\,3.1, there are comparatively short periods with several
obtained spectra between 2018 January 24 and 2018 February 1 as well as in 2023
April/May. For the bulk of the measured maser features, i.e. in particular for
those which are strong enough to be detected at individual epochs with high
signal-to-noise ratios, we do not see any significant changes within the relatively 
short time intervals that cannot be explained by pointing and/or calibration errors. 
Significant differences in lineshapes are not obvious. This also holds for the 
(5,4) line during 2019, Dec. 27 -- 29 (Fig.~\href{zenodo.org/records/15746097}{A.5}) 
as well as for the (9,6) line data with coarse spectral resolution 
(Figs.~\href{zenodo.org/records/15746097}{A.29} and \href{zenodo.org/records/15746097}{A.30}), 
even though this line is known to show the fastest variability (e.g. Henkel et al.
2013). For the (9,6) and (10,7) lines we have an additional  large number of 
monitoring epochs with high velocity resolution, encompassing 2022 and 2023 
(Figs.\,\href{zenodo.org/recoords/15746097}{A.31}, \href{zenodo.org/records/15746097}{A.32}, 
\href{zenodo.org/records/15746097}{A.39} and \href{zenodo.org/records/15746097}{A.40}). 
Here we find slight differences in the (9,6) inversion line between 2013 March 17 and 
May 9 near $V_{\rm LSR}$ = 56\,km\,s$^{-1}$, as well as between 2022 May 4 and June 17 
near 59\,km\,s$^{-1}$. Indications for polarized line components as reported by
Henkel et al. (2013) are not found. There are also small but notable differences 
between 2022 November 25 and 2023 February 7. The high spectral resolution data from 
the (10,7) line do not show such changes.

\subsubsection{Additional ammonia features}

There are a few spectral features that do not fully match the general trends outlined 
above. In the (5,4), (7,6), (7,7), (8,6) and (11,9) transitions, two neighboring 
features are occasionally detected. Earliest evidence for it is seen in the (7,7) 
lines in 2012 and 2013, where the 45\,km\,s$^{-1}$ component is accompanied by a
$\approx$50\,km\,s$^{-1}$ feature (Fig.\,\href{zenodo.org/records/15746097}{A.23}). 
The latter component is getting weaker and is absent in the most recent (7,7) spectra 
(Figs.\,\ref{2023c} and \href{zenodo.org/records/15746097}{A.24}). The low spectral 
resolution profiles of the (5,4) line are not showing convincingly two maser features 
(Figs.~\href{zenodo.org/records/15746097}{A.4} and \href{zenodo.org/records/15746097}{A.5}). 
However, the high resolution spectra taken in 2019 and 2020 are clearly separating the 
components (Fig.~\href{zenodo.org/records/15746097}{A.6}). The (7,6) line also reveals 
both components, but only in 2023 (Fig.\,\href{zenodo.org/records/15746097}{A.22}).  
Particularly clear is the situation in the case of the (8,6) line, where a double 
feature is also seen in 2023, at $\approx$50 and 54\,km\,s$^{-1}$. Earlier, in 2018, 
a double feature at 48 and 54\,km\,s$^{-1}$ is observed. Finally, at the end of January 
2018, a tentative double feature is observed in the (11,9) line. The velocities 
agree with those of the double feature of the (8,6) line observed during the same time.

\subsection{SiO and NH$_3$-VIB}

\begin{figure}[t]
\vspace{-1.3cm}
\resizebox{24.0cm}{!}{\rotatebox[origin=br]{-90}{\includegraphics{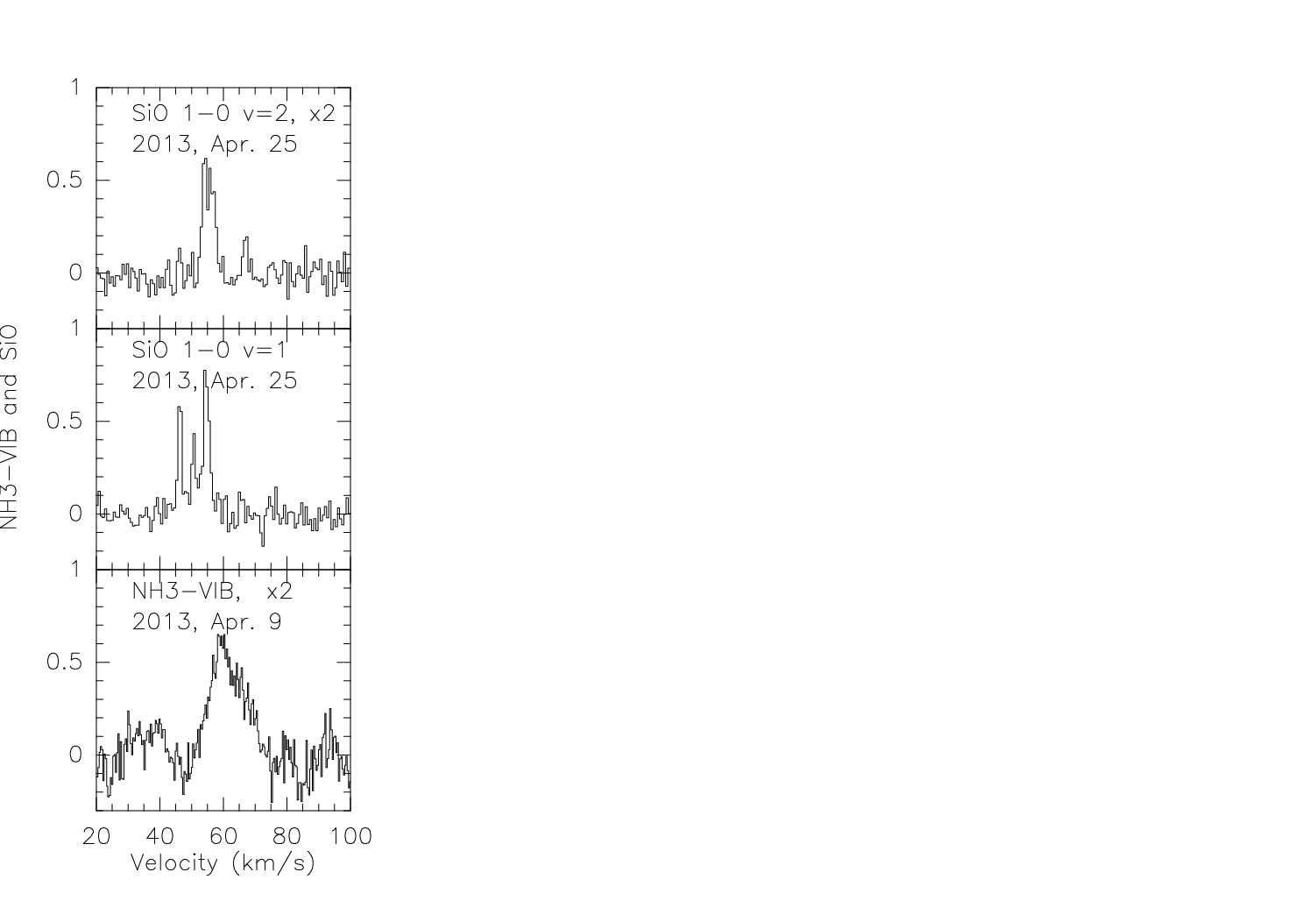}}}
\vspace{-1.3cm}
\caption{Top panel: SiO $J$ = 1$\rightarrow$0 $v$ = 2 on a Jy scale. Middle panel:
SiO $J$ = 1$\rightarrow$0 $v$ = 1 on a Jy scale. Bottom panel: NH$_3$-VIB on a $T_{\rm mb}$
scale. Adopted frequencies are 42.82049, 43.12203, and 466.2452\,GHz from the CDMS
catalogue (M{\"u}ller et al. 2001, 2005; Endres et al. 2016). Channel spacings are 
0.68, 0.68 and 0.39\,km\,s$^{-1}$ from top to bottom. In case of the Gaussian fits outlined 
in Table~B.20 of Appendix B, unsmoothed spectral profiles with velocity spacings 
of 0.17\,km\,s$^{-1}$ have been used for the SiO data.}
\label{nh3vibsio}
\end{figure}

The lines of vibrationally excited ammonia and the ground rotational SiO transitions 
in their first and second vibrationally excited states (Fig.\,\ref{nh3vibsio}) can 
be compared with the K-band ammonia features also measured in 2013 April. In the 
SiO $v$ = 2 line, there is a narrow $<$1\,km\,s$^{-1}$ wide feature at 54.1\,km\,s$^{-1}$, 
which is offset by a few hundred m\,s$^{-1}$ from our so-called 54\,km\,s$^{-1}$ maser 
component from ammonia, i.e. those seen in the (7,4) and (8,5) transitions 
(Table~B.1). We should note that the levels giving rise to the $v$=2 
emission are located at $\approx$3520\,K above the ground state and thus far 
above those of the K-band ammonia lines discussed throughout this paper. Therefore 
the SiO $v$=1 line with $\approx$1770\,K is closer in excitation. And here we find at 
54.6\,km\,s$^{-1}$ a broader component ($\approx$2\,km\,s$^{-1}$), which matches the 
corresponding ammonia feature in a better way. Note, however, that according to Goddi 
et al. (2015), the NH$_3$ and SiO masers are offset by 0\ffas65.

The (0,0) $\rightarrow$ (1,0) transition from the $v_2$ = 1 state of ammonia, the lowest
vibrationally excited level of NH$_3$, is located about 1400\,K above ground and thus 
comparable in excitation to SiO $v$ = 1 or the previously measured ($J,K$) = (12,12) 
line of ammonia in the ground vibrational state (e.g. Henkel et al. 2013). The above
mentioned $v_2$ = 1 ammonia line near 466\,GHz is the equivalent to the ground state
$v_2$ = 0, (1,0) $\rightarrow$ (0,0) line and shows, as the K = 0 ladder in the 
vibrational ground state, no inversion doubling. The profile shown in Fig.\,\ref{nh3vibsio} 
(lowest panel) is not dominated by distinct narrow spectral features but by a single 
wide component covering approximately 20\,km\,s$^{-1}$ and peaking right at the systemic 
velocity of the source. Thus we are not able to connect the line profile to specific 
K-band $\lambda$ $\approx$1.3\,cm ground state maser lines. Instead, it covers 
the velocity range encompassed by the ground state (9,6) line. 

We apply eq.\,(1) of Schilke et al. (1992), i.e.
\begin{equation}
  N_u = 0.01286 \times\ {\rm k_B} \left(\mu^2 \nu_{\rm ul}\right)^{-1} \ \int{T_{\rm mb}\,\,{\rm d}v},
\end{equation}
with k$_{\rm B}$ denoting the Boltzmann constant, $\nu_{\rm ul}$ being the frequency and $\mu$
representing the dipole moment, 1.25\,Debye. Taking the integrated intensity from 
Table~B.20, we thus obtain for the upper state of the observed transition,
the (0,0) level, $N_{\rm u}$ $\approx$ 1.3$\times$10$^{12}$\,cm$^{-2}$. With the partition
function for $T_{\rm ex}$ = 300\,K, 
\begin{equation}
  Q = \Sigma{g_i\ e^{-E/{\rm k_B} T_{\rm rot}}} \approx\ 580
\end{equation}
(see footnote\footnote{https://splatalogue.online/\#/home}), and a statistical weight of 2 because the 
line is an ortho-NH$_3$ transition, this leads to a total population of about $N(v_2 = 1)$ 
$\approx$ 3.8 $\times$ 10$^{14}$\,cm$^{-2}$ inside the 13\arcsec\ sized beam of the APEX telescope.
Realistically, however, the emission must arise from a much smaller region. Mauersberger
et al. (1987) report a source size of 1\ffas26 for the quasi-thermal (7,7) line. Goddi et al.
(2015) provide consistent results for some other highly excited metastable ($J$ = $K$) lines.
Adopting an angular size of order 1\arcsec\ for the $v_2$ = 1, (0,0) $\rightarrow$ (1,0) 
transition, we thus obtain a column density of $N$($v_2$=1,1\arcsec) $\approx$ 6.4 $\times$ 
10$^{16}$\,cm$^{-2}$. For the ground vibrational level, Mauersberger et al. (1987) find 
$N$($v_2$=0) $\approx$2 $\times$ 10$^{19}$\,cm$^{-2}$ for a 1\arcsec\ beam. Aiming at the 
vibrational temperature, we then find with
\begin{equation}
  T_{\rm vib}  =  -\frac{E_{10}/{\rm k_B}}{\rm{ln}({N_1/N_0})}
\end{equation}
$T_{\rm vib}$ $\approx$ 250\,K. This is very similar to the rotational temperature obtained
by Mauersberger et al. (1987) and can be taken as an indication, that the $v_2$=1 line
is not highly opaque. As a cautionary remark, however, we should also note that the submillimeter
continuum at the frequency of the line, $\approx$466\,GHz, may be optically thick, thus 
contaminating comparisons between the $v_2$ = 1 and the ground vibrational state which is
studied at likely more transparent radio frequencies.

\section{Discussion}

\subsection{Maser line saturation}

For many NH$_3$ inversion transitions the line widths of the maser profiles remained 
an unknown quantity. This was mainly related to the spectral resolution of the data 
presented by Henkel et al. (2013). Their channel widths were of order of 1\,km\,s$^{-1}$ 
or higher. The data presented here offer higher spectral resolution profiles, providing 
accurate line widths for several of the measured maser features (see Tables~B.1 -- B.6).

\begin{figure}[t]
\vspace{0.0cm}
\resizebox{9.3cm}{!}{\includegraphics{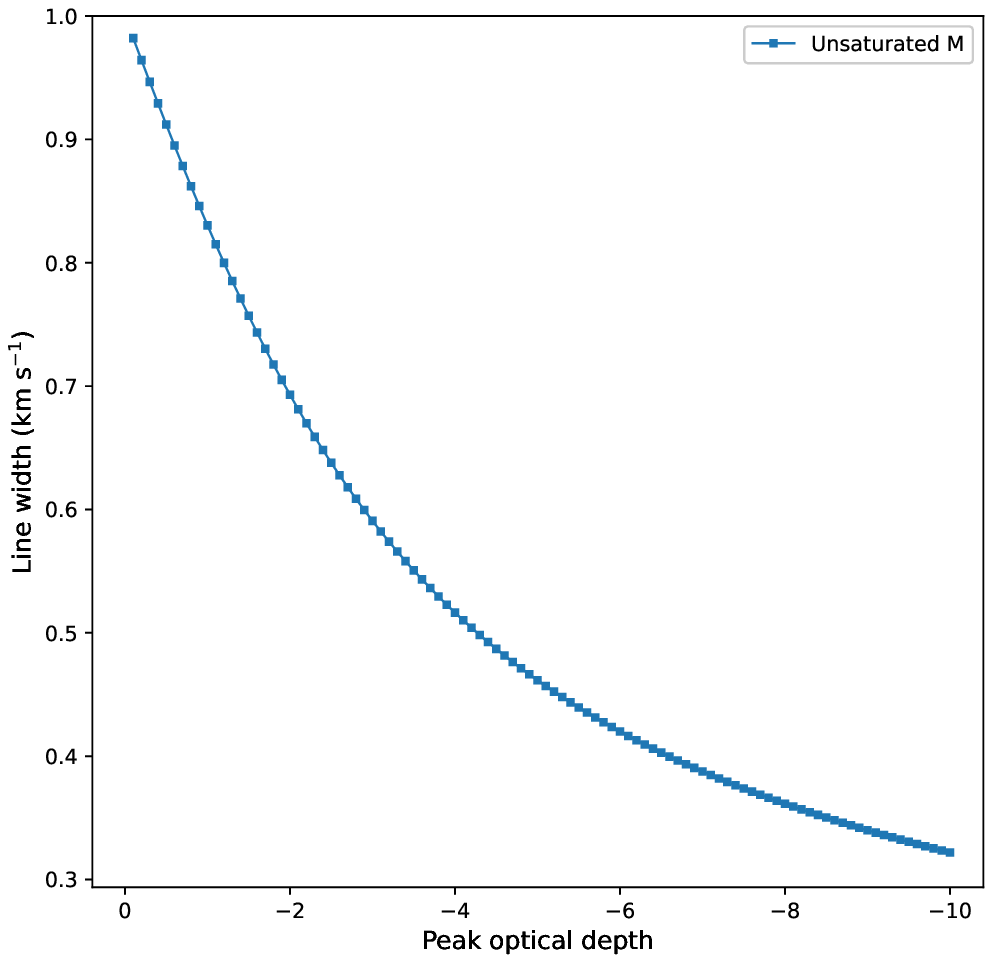}}
\vspace{-0.6cm}
\caption{Narrowing of an unsaturated maser feature with a full width to half maximum of 
1\,km\,s$^{-1}$ as a function of peak optical depth, based on the Rayleigh-Jeans 
approximation of the equation of radiative transfer, $T_{\rm line}$ $\propto$ 
(e$^{-\rm \tau}$ -- 1) and a Gaussian distribution of opacity $\tau$ as a function of 
velocity.} 
\label{maserwidth}
\end{figure}

Line widths well below 1\,km\,s$^{-1}$ could be convincingly determined in case of 
the (5,3), (5,4), (6,3), (7,4), (8,5), (9,7) and (10,7) transitions. In case of the 
45\,km\,s$^{-1}$ features we find full width to half maximum (FWHM) line widths which 
are all around 1\,km\,s$^{-1}$. For the 54\,km\,s maser components line widths are 
in the range of 0.3 -- 0.6\,km\,s$^{-1}$ in the (7,4) (8,5), (9,7) and (10,7) 
transitions, but only $\approx$0.1\,km\,s$^{-1}$ in the (5,4) transition (see 
Fig.~\href{zenodo.org/records/15746097}{A.6}). This is possible, because the main 
hyperfine components of the (5,4) transition only cover about 0.015\,km\,s$^{-1}$, 
a situation similar to that of the (10,7) line as discussed later in this section. 
Finally, for the 57\,km\,s$^{-1}$ features, line widths range between 0.4 and 
0.8\,km\,s$^{-1}$ in the (5,3), (5,4), and (6,3) transitions. The similarity of line widths 
of each of the individual kinematical features (with the notable exception of the
ultra-narrow 54\,km\,s$^{-1}$ (5,4) feature) may support our assumption that many of 
the lines arise inside the same volume of gas.

With the equation of radiative transfer in the Rayleigh-Jeans limit, we obtain
\begin{equation}
  T_{\rm line} \propto \ {\rm e}^{-\tau} - 1, 
\end{equation}
adopting an excitation temperature $T_{\rm ex}$ $<$ 0 as is required for maser emission.
Further assuming a Gaussian distribution of the optical depth as a function of velocity, 
\begin{equation}
  \tau \ = \tau_0 \times\ {\rm e}^{-4\,{\rm ln}2 [(V-V_0)/\Delta V]^2},
\end{equation}
we then get for the full width to half maximum line width
\begin{equation}
  \Delta V_{\rm maser} = 2 \Delta V \left[ \frac{-{\rm ln}(-\tau_0^{-1} \times\ {\rm ln}(1+\,T_{\rm line}/2))}{4\,{\rm ln}2} \right]^{1/2}.
\end{equation}
Here, $T_{\rm line}$ is the peak line temperature at optical depth $\tau_0$ and $\Delta V$ 
is the full width to half maximum line width in case of non-inverted level populations. 
As a result, the maser line becomes, with rising absolute opacity, narrower than it would 
be under quasi-thermal conditions with positive excitation temperature and opacity. This 
is visualized in Fig.\,\ref{maserwidth}.

\begin{figure}[t]
\hspace{-0.2cm}
\resizebox{8.3cm}{!}{\includegraphics{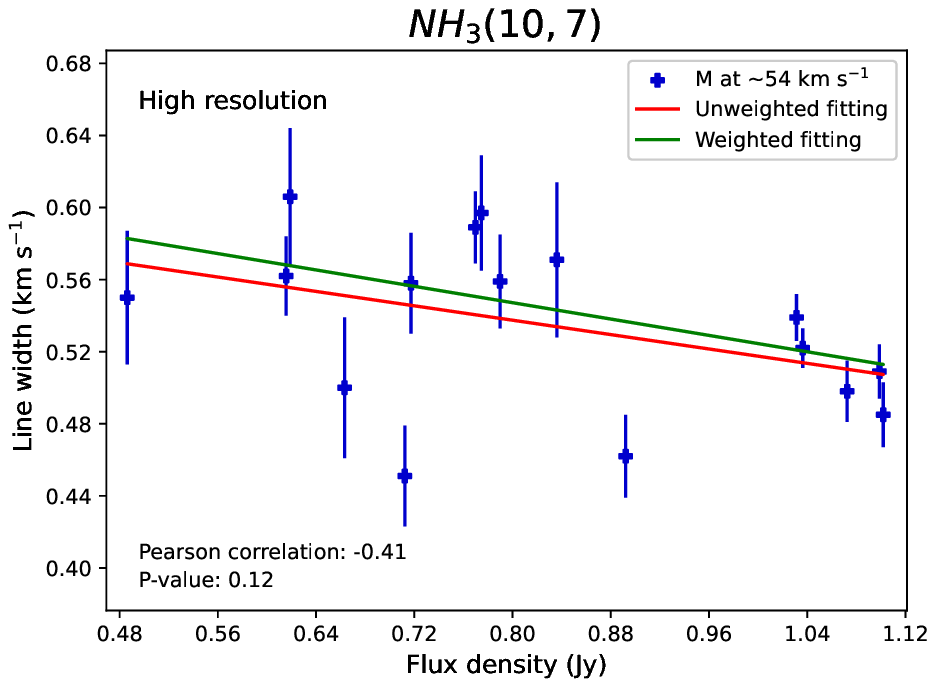}}
\vspace{-0.3cm}
\caption{NH$_3$ (10,7) line widths as a function of line intensities. For the data,
see Table~B.6. The green and red lines represent linear fits to the weighted and 
unweighted data, repectively.}
\label{107a_variability}
\end{figure}

By far the best data allowing for a comparative analysis of line width versus line intensity
are provided by our high spectral resolution monitoring of the (10,7) inversion line
in 2022 and 2023 (see Figs.\,\href{zenodo.org/records/15746097}{A.39} and 
\href{zenodo.org/records/15746097}{A.40} and Table~B.6). Fig.\,\ref{107a_variability} 
shows the corresponding plot. What we find is a range in flux density by a factor of 
two accompanied by a change in line width of $\approx$14\%. As expected, higher flux 
densities are leading to lower line widths. Connecting these observational data with 
the calculated changes shown in Fig.\,\ref{maserwidth}, we find a doubling of the 
flux density (assuming that the excitation temperature of the line does not vary) 
and a decrease in line width by 14\% with opacities between \hbox{--0.7} and 
\hbox{--1.6}. Note that this is more a qualitative than a quantitative estimate, mainly 
because of the only moderate change in line width. Not only that the excitation 
temperature might be changing, but there may also be small calibration errors even 
though the measurements were obtained with the same front- and backends. Furthermore,
line intensities also depend on the pointing accuracy of the telescope. However, pointing 
errors $<$10$^{\prime\prime}$ (Sect. 2.1) do not allow for flux density variations by a 
factor of two. Pointing errors of 3, 5 and 7$^{\prime\prime}$, adopting a point-like maser 
source, only lead to flux density reductions of 1\%, 3\%, and 5\% inside of a 
50$^{\prime\prime}$ beam. Therefore a scenario of unsturated maser emission with 
an opacity of order --1 appears to be a suitable interpretation of the data obtained. 

We have to note, however, that the considerations outlined above are only valid, if
the strongest hyperfine components of the (10,7) transition are not encompassing 
a velocity range that is significant with respect to the observed line width. For 
unsaturated masers it is helpful that the stonger hfs features are amplified more 
efficiently than the weaker ones, enhancing the discrepancy between weaker and 
stronger features (see, e.g., Agafonova et al. 2024). However, in case of the (10,7) 
line the situation is clear. Using the JPL catalog\footnote{
http://spec.jpl.nasa.gov/ftp/pub/catalog/v4/c017002.cat}, we find six centrally
located components, each of them stronger by at least two orders of magnitude 
than the ten individual hfs components $\ga$3\,km\,s$^{-1}$ on each side.
These six strong hfs features only cover a velocity range $\approx$0.01\,km\,s$^{-1}$,
i.e. their velocity coverage is indeed negligible with respect to the measured 
linewidth.

\subsection{Acceleration of the $V_{\rm LSR}$ $\approx$ 45\,km\,s$^{-1}$ component}

Henkel et al. (2013) reported that the lowest velocity maser component, their
"45\,km\,s$^{-1}$" feature, showed a secular drift of $\approx$0.2\,km\,s$^{-1}$\,yr$^{-1}$
towards higher velocities, presumably due to an accelerated outflow also seen in SiO
(Eisner et al. 2002). The monitoring was performed between 1996 and 2012. Our more
recent data encompassing the following decade reveal a more complex picture.

\begin{figure}[t]
\vspace{0.0cm}
\resizebox{8.3cm}{!}{\includegraphics{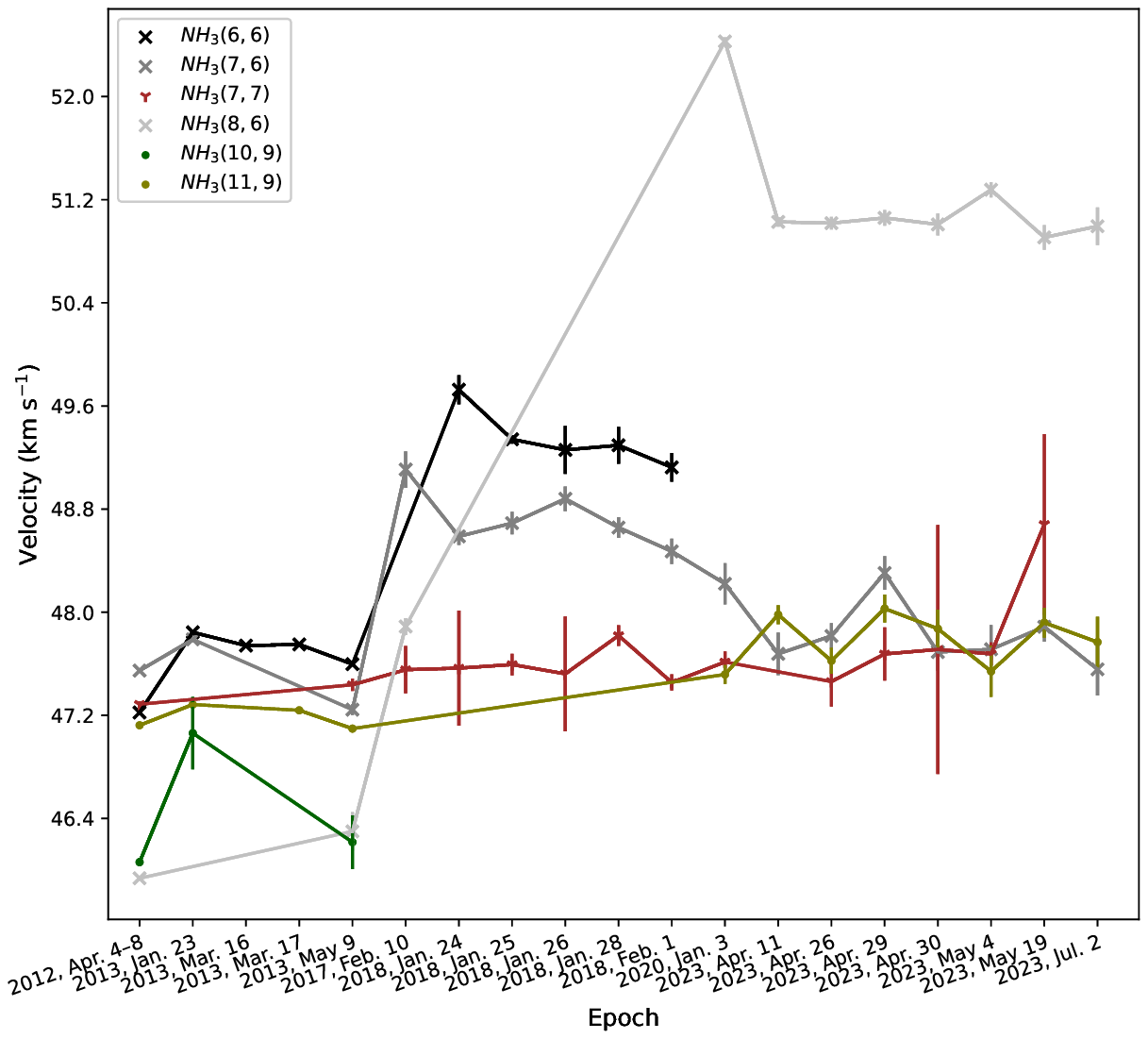}}
\vspace{-0.3cm}
\caption{Radial velocity of the so-called $V_{\rm LSR}$ = 45\,km\,s$^{-1}$ component
as a function of time for various NH$_3$ inversion lines.} 
\label{velocitydrift}
\end{figure}

From Fig.\,\ref{velocitydrift} we can see that there is an overall trend towards 
48\,km\,s$^{-1}$, as seen in the (7,7), (7,6) and (11,9) transitions. However, 
this infers a much lower drift velocity than observed before, a drift at the order 
of 0.07\,km\,s$^{-1}$. In addition, there are other lines which show quite a 
different behavior. The (6,6) line drifted upwards between 2013 and 2017 by 
about 2\,km\,s$^{-1}$ or created a new feature while the one at lower velocity
faded. The difference in velocities is even more pronounced in the (8,6) 
line, where the velocity is rising from $\approx$46\,km\,s$^{-1}$ to 51--52\,km\,s$^{-1}$
between 2013 and 2023. In view of the lines, where the shift in velocity
seems to be drastically enhanced during the last decade, we find it likely 
that the former maser feature has disappeared to be replaced by another one 
at higher velocity.

\subsection{The ($J,K$) = (9,6) as a special case}

As mentioned, the (9,6) line is the fastest varying and strongest
ammonia inversion line (see Figs.\,\href{zenodo.org/records/15746097}{A.29} 
-- \href{zenodo.org/records/15746097}{A.32}). The main spike is found, 
in agreement with Zhang et al. (2022), at velocities near 
54\,km\,s$^{-1}$, slightly above this value in 2013 to decrease
slightly below this value in 2022 and 2023 (see also Sect\,3.4.1). This 
is not seen in the 54\,km\,s$^{-1}$ features of other lines observed with 
high spectral resolution. The components near 60\,km\,s$^{-1}$ did also shift 
to slightly lower velocities. Noteworthy is in particular the emission 
below 50\,km\,s$^{-1}$, which can be found in the high spectral resolution 
plots of Figs.\,\href{zenodo.org/records/15746097}{A.31} and 
\href{zenodo.org/records/1574609}{A.32}, which has to our knowledge not 
been reported before. The most recent spectra show components near 49 and 
even 47\,km\,s$^{-1}$. The latter emission corresponds to that seen also 
in other lines as shown in Fig.\,\ref{velocitydrift}, leading to a velocity 
component observed in four consecutive inversion lines along the same K-ladder,
i.e. the (6,6), (7,6), (8,6) and (9,6) transitions. We have to note,
however, that the corresponding components of the (6,6) and (8,6) 
lines have disappeared at these epochs.

\subsection{Para- and ortho-NH$_3$ masers}

The recent survey by Yan et al. (2024) revealed that in most massive star formation
regions with NH$_3$ masers, either solely para- or ortho-NH$_3$ maser transitions
could be identified. Only one of their 14 newly discovered NH$_3$ maser
sources revealed inverted populations in both NH$_3$ maser species. Clearly,
W51-IRS2 reveals maser lines from both ammonia species. However, this may
only be due to the superposition of different maser hotspots in our single-dish
beam (see, e.g., our Fig.\,2 and fig.\,5 in Zhang et al. 2024). For the so-called 
45\,km\,s$^{-1}$ components we find inverted populatiosn in the (6,6), (7,6), 
(7,7), (8,6), (10,9) and (11,9) spectra (see Sect. 3.4). Furthermore, Henkel 
et al. (2013) reported corresponding features also in the (9,9) and (12,12) transitions. 
So this component reveals almost exclusively ortho-NH$_3$ masers with the notable 
exception of the (7,7) line.  The 54\,km\,s$^{-1}$ feature is seen in the (6,2), 
(7,3), (7,4), (7,5), (8,5), (9,7), (9,8), and (10,7) lines (Sect. 3.4). So this 
is an opposite case to that of the 45\,km\,s$^{-1}$ component. All lines belong 
to the para-NH$_3$ species except the (7,3) transition. 57\,km\,s$^{-1}$ features 
are seen in the (5,2), (5,3), (5,4), (6,3) and (6,4) line. Here both para- and 
ortho-NH$_3$ are contributing.

The preferences for either para- or ortho-NH$_3$ might be explained by column density 
effects, because ortho-NH$_3$ levels are characterized by twice the statistical 
weights of their para-counterparts, thus requiring lower column densities to reach 
optical depths yielding noticeable exponential maser amplification in cases of 
unsaturated maser emission (see Sect. 3.4.2). However, such an explanation may allow 
for deviations from the rule, since a factor of two in the statistical weights 
is not very much.

\subsection{Comparison with other data}

Following e.g. Goddi et al. (2015) and Zhang et al. (2024), the warm molecular gas 
in W51-IRS2 shows a structure that is elongated along an east-west axis, including 
W51-N4, W51-d2, W51-N2 and W51-North from west to east, encompassing about 
6$^{\prime\prime}$ (0.15\,pc) at the southern edge of the more extended H{\sc ii} 
region W51d (Fig.~\ref{goddi}). This includes the so-called Lacy-jet (Lacy et al. 
2007), representing ionized gas, which originates in W51-N4, i.e. in the west, 
and extends eastwards across the other above mentioned sub-sources (see, e.g., 
the upper left panel of fig.\,8 in Zhang et al. 2024).

With respect to the discussion in Sect.\,4.4, we thus may ask whether the 57\,km\,s$^{-1}$
maser features, representing both para- and ortho-NH$_3$ transitions, arise from
different regions inside W51-IRS2. Unfortunately, high angular resolution NH$_3$ 
data provide so far only one maser line with emission near 57\,km\,s$^{-1}$. It 
is the ($J$,$K$) = (6,3) ortho-NH$_3$ transition, reported by Zhang et al. (2024), 
and observed towards W51d2 (their fig.\,3). While we have no accurate position for 
a corresponding para-NH$_3$ line, it may be possible that such maser lines might 
arise from W51-North, located about 3\ffas5 eastwards, because this is the region
where most of the W51-IRS2-masers originate (e.g. Goddi et al. 2015; Zhang et al. 
2023, 2024). For example, the  ``45\,km\,s$^{-1}$'' masers are observed at this 
position.

\begin{figure}[t]
\vspace{0.0cm}
\resizebox{8.4cm}{!}{\rotatebox[origin=br]{0}{\includegraphics{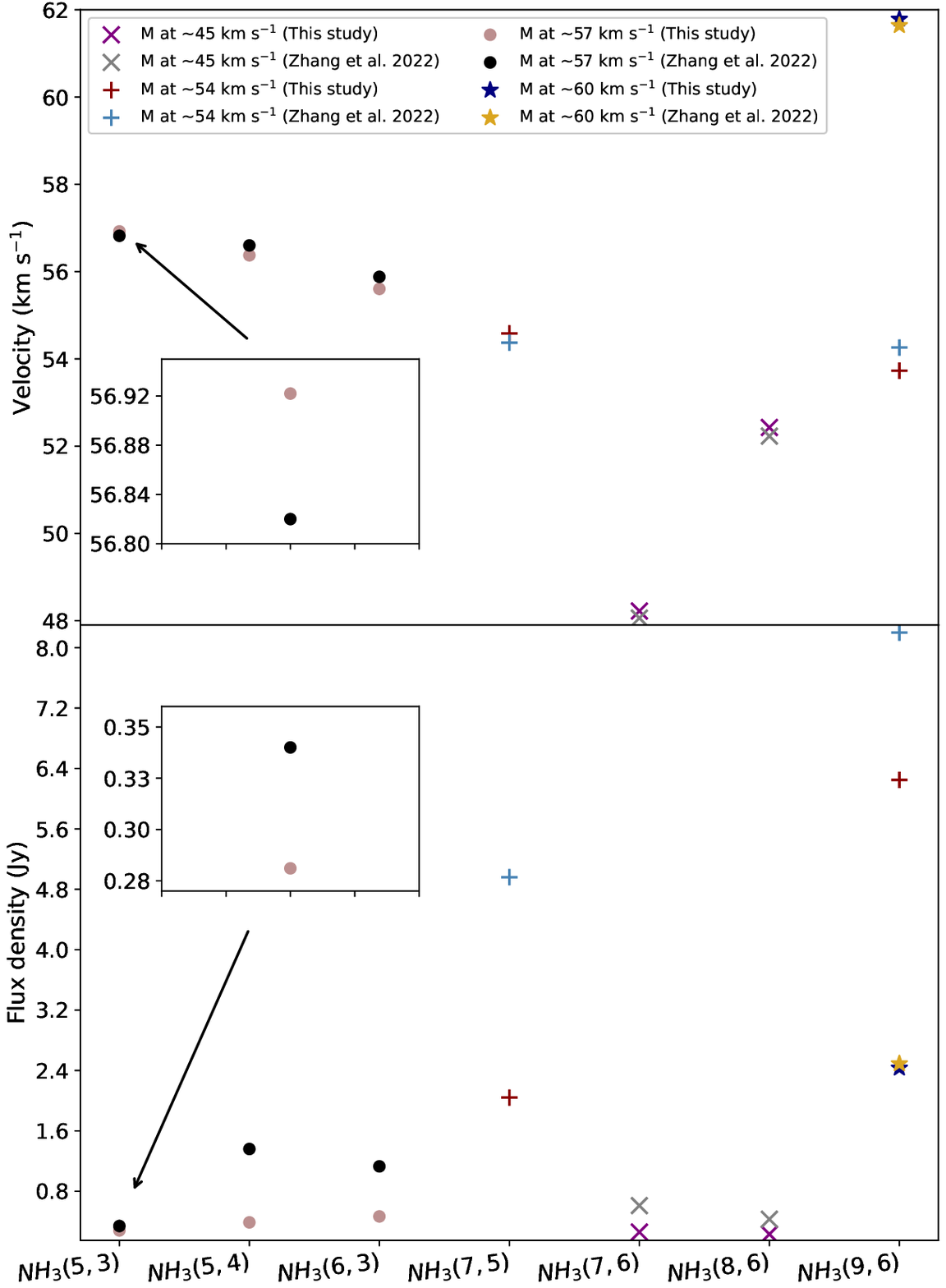}}}
\vspace{0.0cm}
\caption{All of the NH$_3$ maser features detected by Zhang et al. (2022) and this
study in January 2020. Note that our data were taken on January 3, while those of
Zhang et al. (2022) were obtained on January 8.}
\label{flux_vs_zhang}
\end{figure}

A monitoring program during 2020, January to April, involving four epochs and tracing 
the molecular species ammonia, water, and methanol was presented by Zhang et al. (2022). 
They used the Shanghai 65\,m Tianma Radio Telescope (TMRT) with beam sizes of order 
50$''$ to 60$''$ (see Fig.~\ref{ginsburg}) and 3$\sigma$ noise levels of $\approx$0.5\,Jy 
in a single channel covering 0.043\,km\,s$^{-1}$. The epochs include early January, mid of 
March and early April and report a burst of line emission in some specific maser transitions. 
Fortuitously, the first epoch, January 8, 2020, used by Zhang et al. (2022) near the 
peak of the H$_2$O maser flare is quite close to January 3, 2020, when we also covered 
the entire 18 -- 26\,GHz K-band spectrum (for a visualization, see Fig.\,\ref{flux_vs_zhang}). 
Zhang et al. (2022) reported rapidly increasing and decreasing line fluxes for several 
of our maser lines, but also for the quasi-thermal ones. In addition, they reported 
absorption features at $\approx$60\,km\,s$^{-1}$ on April 7. 

Comparing our data from January 3, 2020, with those by Zhang et al. (2022) taken
five days later, we can check some of their results in the light of our spectra. 
Zhang et al. (2022) reported a tentative ($J$,$K$) = (8,7) maser feature that is 
not confirmed by our data. While we see, like Zhang et al. (2022), more than one 
velocity feature in the (5,4), (7,6) and (9,6) maser lines, we do not find 
(unlike Zhang et al. 2022) more than a single non-thermal velocity component in 
the (7,5) transition. Confirming Zhang et al. (2022), there are also single maser 
components in the (5,3), (7,3), (8,6), and (10,7) lines on 2020 January 3, with 
our (8,6) line data revealing two velocity components at earlier and later epochs. 
The (9,7) and (9,8) lines are not detected by us during the critical time interval 
discussed here. Our decade-long monitoring observations never revealed any absorption 
line. Thus we do not find any supporting evidence for the (4,1), (4,2), (5,2), (5,3),
(6,4), (8,6) and (9,6) absorption features reported by Zhang et al. (2022) 
near $V_{\rm LSR}$ = 60\,km\,s$^{-1}$ as we can also not confirm variability
in the quasi-thermal lines (see Sect.\,3.1).  

While there is good agreement with the absence of (6,6) maser emission and only weak 
maser emission in the (10,9) line in January 2020, another potential discrepancy 
is related to the (7,7) transition. The non-detection of the (7,7) maser by Zhang 
et al. (2022) is not compatible with our $S_{\rm \nu}$ $\approx$0.3\,Jy detection 
only five days prior to their first epoch unless the maser is varying on very 
rapid timescales, which is not apparent in our longer term monitoring data.

Following the peak flux densities given in Zhang et al.'s (2022) fig.\,5 and our
data from an epoch only five days earlier (2020, January 3), we find good agreement 
in the (5,3), (6,2) and (7,3) transitions, moderate agreement in the (8,6) line, 
and clear disagreement in the (5,4), (6,3), (7,5) and (7,6) maser features.
We suspect that these differences are not predominantly caused by maser 
variability in view of our short time scale analysis provided in Sect.\,3.4.3.
Thus we are viewing the data of Zhang et al. (2022) with some skepticism and 
suggest that variability is not as extreme and fast as suggested in their 
analysis from spectra taken in early 2020.

\subsection{Heat waves and flaring events}

\begin{figure}[t]
\vspace{0.0cm}
\resizebox{8.4cm}{!}{\rotatebox[origin=br]{0}{\includegraphics{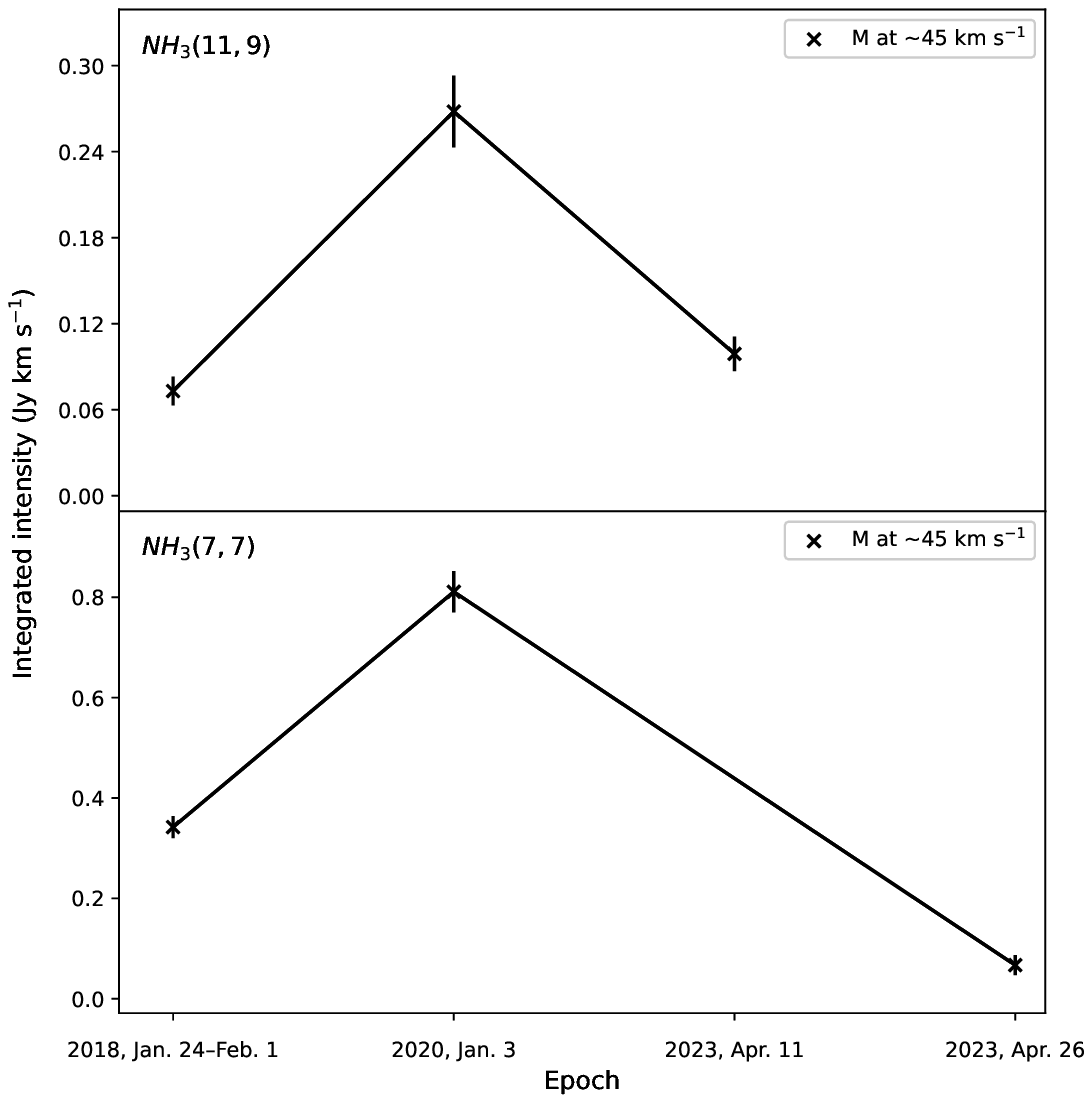}}}
\vspace{0.0cm}
\caption{Variability of the integrated intensities of the (7,7) and (11,9) lines 
between 2018 and 2023.}
\label{77-119}
\end{figure}

As already mentioned in Sect.\,4.5, Zhang et al. (2022, 2023) reported luminosity 
outbursts in H$_2$O, NH$_3$ and CH$_3$OH maser lines in early 2020. With our
data from 2020, January 3, we can check whether our maser lines are flaring 
with respect to early 2018, i.e. two years earlier, and April 2023, about 
three years afterwards. Our results turn out to be mostly ``disappointing''.
Among the maser features with sufficient signal-to-noise ratios to be 
analyzed in this context, the (5,3), (5,4), (6,3), (7,3), (7,4), (7,5), (7,6), 
(8,6), (9,6) and (10,7) transitions show no significant increases in flux density
in early 2020. Only the (7,7) and (11,9) lines show such an effect (Fig.~\ref{77-119}). 
Integrated (7,7) maser line intensities in units of Jy\,km\,s$^{-1}$ are 0.342$\pm$0.022 
on 2018, February 1, 0.811$\pm$0.041 on 2020, January 3, and 0.067$\pm$0.020 on 2023, 
April 26 (Table~B.14). For the (11,9) line we find corespondingly
0.073$\pm$0.010 (2018, Jan. 24 to Feb. 1), 0.268$\pm$0.025 (2020, Jan. 3) and
0.099$\pm$0.12 (2023, Apr. 11) in Table~B.19). To conclude, we do 
not see a general flare in the ammonia maser lines in early 2020.

A longer monitoring time span from 2020 till 2023, available for the 22\,GHz 
H$_2$O line, was reported by Volvach et al. (2023). Their data, obtained with 
an angular resolution of 150$^{\prime\prime}$ and thus potentially including 
W51-IRS1 (see Fig.\,\ref{ginsburg}), reveal a strong flare between late 2021 and 
early 2022, peaking approximately in January 2022 (their fig.\,1). Zhang et al. (2023) 
reported that the bulk of this flare likely originates from a compact disk-like 
structure in W51-North, the eastern part of W51-IRS2. The 22\,GHz H$_2$O transition 
connects levels $\approx$640\,K above the ground state which is compatible with 
several of our ammonia masers (see Table~1). We may therefore ask whether we can
find any evidence for this flare also in our NH$_3$ data.

We present observations in early 2022 for the (9,6) and (10,7) lines. These are 
two particularly highly excited NH$_3$ inversion lines connecting levels $>$1000\,K
above the ground state (Table~1). While the peak flux density of the water maser flare 
was about 140\,kJy, which is four orders of magnitude more powerful than the most 
intense maser component we detected in this study, H$_2$O flux densities decrease
from $\approx$20\,kJy in early March 2022 to $\approx$15\,kJy till the end of June 
to then briefly rise to 30\,kJy in early July, decreasing then once more to 15\,kJy, 
rising then during the last months of the year to 20--25\,kJy to then fade to normal 
levels of a few kJy. In our high spectral resolution data of the (9,6) transition 
(Figs.\,\href{zenodo.org/records/15746097}{A.31} and \href{zenodo.org/records/15746097}{A.32}) 
we do not see any correlated variability. This also holds for the (10,7) masers 
presented in Figs.\,\href{zenodo.org/records/15746097}{A.39} and 
\href{zenodo.org/records/15746097}{A.40}. It should be noted here, that the NH$_3$ 
masers in W51-IRS2 may be pumped by infrared radiation (see Sect.\,3.4.2), while 
the H$_2$O masers are believed to be collisionally excited (e.g. Kylafis \& Norman 
1987, 1991). This might lead to delayed H$_2$O maser flare episodes with respect to 
NH$_3$, also depending on the detailed spatial configuration.
 
Zhang et al. (2022) interpreted variability in terms of rather shortlived accretion
episodes, leading to heatwaves and to maser flares in the surrounding energized medium. 
What we have found is a maser flare identified on much larger time scales. With 
respect to the so-called 45\,km\,s$^{-1}$ component, emission was strongest during 
the initial epochs (see Henkel et al.'s 2013 figs.\,1--3 and our Appendix A). 
Originally discovered in the ammonia (9,9) and (12,12) lines of orthe-NH$_3$ in 
November and December 1995, the multilevel ammonia study of Mauersberger et al.
(1987) presenting spectra from 1985 and 1986 does not show any emission at this
velocity. The (9,9) and (12,12) lines showed increased line intensities in late 
August of 2011 (Henkel et al. 2013). The lines studied here with sufficient signal-to-noise 
ratios are the (6,6), (7,6), (7,7), (8,6), (11,9) transitions. What is common to their 
maser features is a slowly advancing decay, which appears not to be monotonic but 
proceeds in distinct steps. The earliest measurements of the (6,6), (7,6), (7,7) masers 
were taken in 2008. For the (8,6) and (11,9) lines, the starting point is early April 
2012. Strongest line emission was seen in 2008, October 25 (Henkel et al. 2013), and 
in 2012 in case of the (6,6) and (7,6) lines, which is  compatible with the (9,9) and 
(12,12) maser lines, which were stonger in 2011 than during their discovery in 1995. 
While the (7,7) transition fades already by a factor of about eight between 2008 and
2012, the (6,6), (7,6) and (8,6) lines only get clearly weaker during 2013. By 2020, the 
(6,6) line is gone. The (7,7) line is not clearly detectable after 2020, while the (8,6)
line stays at a constant level of $\approx$0.25\,Jy. The latter also holds for the (11,9)
transition, only that here the resulting flux densities are below 0.1\,Jy. And the (9,6) 
transition starts to show weak emission at velocities slightly below 50\,km\,s$^{-1}$.

To summarize, the feature may have risen for at least 15 years from 1995 till about 2010,
while it was not yet apparent in 1985/1986 (Mauersberger et al. 1987). Beginning in 2013,
the lines became weaker. Here it is interesting that the line with lowest excitation, the
(6,6) transition, is fading first, while the (8,6) and (11,9) transitions still show the feature
in 2023. During this year, also weak (9,6) emission at $V_{\rm LSR}$ $<$ 50\,km\,s$^{-1}$
is appearing. Instead of a heat wave which expands and successively excites lines of lower
energy above the ground state, here we seem to have the opposite with only the lines with 
highest excitation still being found to be masering. Not knowing exact physical boundary 
conditions nor having detailed radiative transfer calculations involving vibrational 
excitation, we propose the following tentative explanations;

\begin{itemize} 
\item We may assume the presence of two sources, one emitting the maser lines at lower 
levels above the ground state, and the other longer lived source providing emission from
the higher $J$ levels. Since both emit at velocities, which cannot be distinguished,
and since both were appearing during the same, admittedly large time interval, this appears
to be not likely. Furthermore, Goddi et al. (2015) demonstrated that the low velocity (blue) 
(6,6), (7,7) and (9,9) line components originate from the same location. Potential differences 
in position are not larger than a few milli-arcsec ($\approx$20\,AU).

\item The possibility that infrared line overlaps play a significant role is also not
likely. This is due to the fact that only the (9,6) line intensities appear to be exceptional,
thus offering the possibility that this but only this line is affected by such 
overlaps (e.g. Madden et al. 1986). Also, here we have several lines with lower and 
higher $J$ quantum numbers.  Similar line overlap effects for several of these transitions 
are unlikely. 

\item Considering the time scales involved, rotation of a molecular cloud is yet another 
option (e.g. Gray et al. 2019). When the (6,6) and (7,7) lines would arise farther away from 
the exciting source, then they might leave first the line-of-sight to a compact background 
source they may amplify. Or the maser cones of the (6,6), (7,6) and (7,7) lines are narrower 
than those of the (8,6) and (11,9) transitions.

\item Intuitively, one may think that masers from lines closer to the ground state are always 
emitting from hotspots farther away from the central irradiating source than the higher $J$ ones. 
However, there exists a counter-example: The SiO $v$ = 1 $J$ = 2$\rightarrow$1 masers are occasionally 
observed at larger distances from the central source than the corresponding $J$ = 1$\rightarrow$0 
lines, representing a lower degree of rotational excitation. This holds in case of the SiO maser 
sources in Orion-KL and the late type stellar object IRC+10011 (e.g. Soria-Ruiz et al. 2004; Issaoun 
et al. 2017). If this is also valid for our ammonia masers, then a passing heat wave might quench 
the inner maser lines from lower $J$ states earlier than those from higher $J$ levels farther out. 

\end{itemize}

In case of the other maser features we lack information on the epoch when they formed. This also 
holds for our tentatively detected (5,2) and (6,4) masers at $\approx$57\,km\,s$^{-1}$, which - if 
real - may have been too weak to be detectable in previous studies. Furthermore, the flux densities 
of the masers at $>$50\,km\,s$^{-1}$ do not follow a sufficiently regular pattern that is 
comparable to that of the so-called 45\,km\,s$^{-1}$ feature. Therefore we cannot provide related 
scenarios potentially explaining their variability.

\section{Conclusions}

We have presented data from the numerous detected ammonia masers in W51-IRS2. Some of 
the key findings of this study are:

(1) We present the first tentative discovery of a ($J$,$K$) = (5,2) ammonia maser
    in the interstellar medium. Tentatively, a (6,4) maser is seen for the first 
    time in W51-IRS2, while, also for the first time, (9,6) maser emission is 
    detected at local standard of rest velocities $<$50\,km\,s$^{-1}$ in this source. 

(2) The data taken during the time interval from 2012 to 2023 demonstrate that previously
    determined peak flux densities are roughly confirmed, because most of the maser 
    lines have full width to half maximum line widths $\approx$1\,km\,s$^{-1}$.

(3) The distribution of maser lines in the NH$_3$ level diagram shows strong hints 
    for being pumped by infrared radiation.

(4) Vibrationally excited ammonia is indeed detected, suggesting a vibrational
    excitation temperature, which is compatible with the kinetic temperature
    of $\approx$300\,K.

(5) Analyzing the (10,7) maser features over a time interval of $\approx$15 months,
    a weak correlation between line width and intensity is found, which is consistent
    with unsaturated maser emission and a peak optical depth of order --1.

(6) The previously reported velocity drift of the so-called 45\,km\,s$^{-1}$ 
    maser component has slowed down considerably or is even absent in most 
    recent spectra. Furthermore, the maser lines from this component connecting
    levels $<$800\,K above the ground state disappear during the observed 
    time interval, while the lines with highest excitation, the (8,6), (9,6) and
    (11,9) lines still show features at $<$50\,km\,s$^{-1}$ till 2023. Tentative 
    interpretations are provided. 

(7) For early 2018 and spring 2023, where monitoring took place on almost daily or 
    weekly basis, no sudden and/or strong variability is seen. No absorption line
    is found. A correlation of ammonia maser lines with a strong 22\,GHz H$_2$O 
    flare is not found. The 54\,km\,s$^{-1}$ features show highly inconsistent 
    variability patterns. It is suggested that they may arise from more than one 
    active region in W51-IRS2.

\section*{Data availibility}
The Appendices A (containing 44 figures with the monitored spectra) and B (including
20 tables presenting Gaussian fits to the spectra) are available at
https://zenodo.org/records/15746097 .

\begin{acknowledgements}

We wish to thank the people at Effelsberg for their great support, in particular B. 
Winkel and A. Kraus, when analyzing the 8\,GHz wide spectra from early 2018. We also
profited from discussions with T. Krichbaum and B. Lankhaar. Finally, we wish 
to thank the referee for good and helpful comments further improving the paper. E. Alkhuja 
was supported by King Abdulaziz University and the Cultural Office of the Embassy of the Kingdom 
of Saudi Arabia in Berlin. Related to A. Wootten: The National Radio Astronomy Observatory 
and Green Bank Observatory are facilities of the U.S. National Science Foundation operated 
under cooperative agreement by Associated Universities, Inc. This research has made use of 
NASA's Astrophysical Data System. 

\end{acknowledgements}


\begin{thebibliography}{}
 \bibitem[2024]{xyz}
  Agafonova, I.~I., Bayandina, O.~S., Gong, Y., et al. 2024, MNRAS, 533, 1714
 \bibitem[2007]{xyz}
  Beuther, H., Walsh, A.~J., Thorwirth, S., et al. 2007, A\&A, 466, 989
 \bibitem[2007]{xyz}
  Bisschop, S.~E., J$\oslash$rgensen, J.K., van Dishoeck, E.~F., \& de Wachter, 
  E.~B.~M. 2007, A\&A. 465, 913 
 \bibitem[1988]{xyz}
  Brown, P.~D., Charnley, S.~B., \& Millar, T.~J., 1988, MNRAS, 231, 409
 \bibitem[1993]{xyz}
  Caselli, P., Hasegawa, T.~I., \& Herbst, E. 1993, ApJ, 408, 548
 \bibitem[1992]{xyz}
  Cesaroni, R., Walmsley, C.~M., Churchwell, E. 1992, A\&A 256, 618
 \bibitem[1992]{xyz}
  Charnley, S.~B., Tielens, A.~G.~G.~M., \& Millar, T.~J. 1992, ApJ, 399, L71
 \bibitem[2023]{xyz}
  Demes, S., Lique, F., Loreau, J. \& Faure, A. 2023, MNRAS, 524, 2368
 \bibitem[2002]{xyz}
  Eisner, J.~A., Greenhill, L.~J., Herrnstein, J.~R., Moran, J.~M., \& Menten, K.~M.,
  2002, ApJ, 569, 334
 \bibitem[2016]{xyz}
  Endres, C.~P., Schlemmer, S., Schilke, P., Stutzki, J., \& M{\"u}ller, H.~S.~P. 
  2016, Jou. Mol. Spec. 327, 95
 \bibitem[2013]{xyz}
  Garrod, R.~T., \& Widicus Weaver, S.~L. 2013, ChRv, 113, 8939
 \bibitem[1996]{xyz}
  Gaume, R.~A., Wilson, T.~L., \& Johnston, K.~J. 1996, ApJ, 457, L47
 \bibitem[1982]{xyz}
  Genzel, R., Ho, P.~T.~P., Bieging, J., \& Downes, D. 1982, ApJ, 259, L103
 \bibitem[2016]{xyz}
  Ginsburg, A., Goss, W.~M., Goddi, C., et al. 2016, A\&A, 595, A27
 \bibitem[2015]{xyz}
  Goddi, C., Henkel, C., Zhang, Q., Zapata, L., \& Wilson, T.~L. 2015, A\&A, 573, A109
 \bibitem[2019]{xyz}
  Gray, M.~D., Baggott, J., Westlake, J., \& Etoka, S. 2019, MNRAS 486, 4216
 \bibitem[1983]{xyz}
  Guilloteau, S., Wilson, T.~L., Martin, R.~N., Batrla, W. \& Pauls, T.~A. 1983, A\&A, 124, 322
 \bibitem[2006]{xyz}
  G{\"u}sten, R., Nyman, \AA., Schilke, P., et al. 2006, A\&A, 454, L13
 \bibitem[1987]{xyz}
  Henkel, C., Jacq, T., Mauersberger, R., Menten, K.~M., \& Steppe, H. 1987, A\&A, 188, L1
 \bibitem[2013]{xyz}
  Henkel, C., Wilson, T.~L., Asiri, H., \& Mauersberger, R. 2013, A\&A, 549, A90
 \bibitem[1997]{xyz}
  Howard, E.~M., Koerner, D.~W., \& Pipher, J.~L. 1997, ApJ, 477, 738
 \bibitem[2017]{xyz}
  Issaoun, S., Goddi, C., Matthews, L.~D., et al. 2017, A\&A, 606, A126
 \bibitem[2012]{xyz}
  Klein, B., Hochg{\"u}rtel, S., Kr{\"a}mer, L., et al. 2012, A\&A, 542, L3
 \bibitem[2014]{xyz}
  Klein, T., Ciechanowicz, M., Leinz, C., et al. 2014, IEEE, 4, No. 5, 588
 \bibitem[1987]{xyz}
  Kylafis, N.~D., \& Norman, C. 1987, ApJ, 323, 346
 \bibitem[1991]{xyz}
  Kylafis, N.~D., \& Norman, C. 1991, ApJ, 373, 525
 \bibitem[2023]{xyz}
  Loreau, J., Faure, A., Lique, F., Demes, S., \& Dagdigian, P.~J. 2023, MNRAS, 526, 3213
 \bibitem[1996]{xyz}
  Madden, S.~C., Irvine, W.~M., Mathhews, H.~E., Brown, R.~D., \& Godfrey, P.~D. 1986, ApJ, 300, 79
 \bibitem[1994]{xyz}
  Mangum, J.~G., \& Wootten, A. 1994, ApJ, 428, L33
 \bibitem[1986]{xyz}
  Mauersberger, R., Henkel, C., Wilson, T.~L., \& Walmsley, C.~M. 1986a, A\&A, 162, 199
 \bibitem[1986]{xyz}
  Mauersberger, R., Wilson, T.~L., \& Henkel, C. 1986b, A\&A, 160, L13
 \bibitem[1987]{xyz}
  Mauersberger, R., Henkel, C., \& Wilson, T.~L. 1987, A\&A, 173, 352
 \bibitem[1988]{xyz}
  Mauersberger, R., Wilson, T.L., \& Henkel, C. 1988, A\&A, 201, 123
 \bibitem[2020]{xyz}
  Mei, Y., Chen, X., Shen, Z.-Q., \& Li, B. 2020, ApJ, 898, 157
 \bibitem[2018]{xyz}
  Mills, E.~A.~C., Ginsburg, A., Clements, A.~R., et al. 2018, ApJ, 869, L14
 \bibitem[2001]{xtz}
  M{\"u}ller, H.~S.~P., Thorwirth, S., Roth, D.~A., \& Winnewisser, G. 2001, A\&A, 370, L49
 \bibitem[2005]{xyz}
  M{\"u}ller, H.~S.~P., Schl{\"o}der, F., Stutzki, J., \& Winnewisser, G. 
  2005, Jou. Mol. Structure, 742, 215
 \bibitem[1994]{xyz}
  Ott, M., Witzel, A., Quirrenbach, A., et al. 1994, A\&A, 284, 331
 \bibitem[1998]{xyz}
  Pickett, H.~M., Poynter, R.~L., Cohen, E.~A., et al. 1998, JQSRT, 60, 8
 \bibitem[2010]{xyz}
  Sato, M., Reid, M.~J, Brunthaler, A., \& Menten, K.~M. 2010, ApJ, 720, 1055
 \bibitem[1991]{xyz}
  Schilke, P., Walmsley, C.~M., \& Mauersberger, R. 1991, A\&A, 247, 516
 \bibitem[1992]{xyz}
  Schilke, P, G{\"u}sten, R., Schulz, A., Serabyn, E., \& Walmsley, C.~M. 1992, A\&A, 261, L5
 \bibitem[2004]{xyz}
  Soria-Ruiz, R., Alcolea, J., Colomer, F., et al. 2004, A\&A, 426, 131
 \bibitem[2023]{xyz}
  Volvach, A. E., Volvach, L. N., \& Larionov, M. G. 2023, ApJ, 955, 10 
 \bibitem[2005]{xyz}
  Wakelam, V., Selsis, F., Herbst, E., \& Caselli, P. 2005, A\&A, 444, 883
 \bibitem[1983]{xyz}
  Walmsley, C.~M., \& Ungerechts, H. 1983, A\&A, 122, 164
 \bibitem[1987]{xyz}
  Walmsley, C.~M., Hermsen, W., Henkel, C., Mauersberger, R., \& Wilson, T.~L. 1987, A\&A, 172, 311
 \bibitem[2007]{xyz}
  Walsh, A.~J., Longmore, S.~N., Thorwirth, S., Urquhart, J.~S., \& Purcell, C.~R. 2007, MNRAS, 382, L35
 \bibitem[1982]{xyz}
  Wilson, T.L., Batrla, W., \& Pauls, T. 1982, A\&A, 110, L20
 \bibitem[2006]{xyz}
  Wilson, T.~L., Henkel, C., \& H{\"u}ttemeister, S. 2006, A\&A, 460, 533
 \bibitem[2022a]{xyz}
  Yan, Y.~T., Henkel, C., Menten, K.~M., et al. 2022a, A\&A, 659, A5
 \bibitem[2022b]{xyz}
  Yan, Y.~T., Henkel, C., Menten, K.~M., et al. 2022b, A\&A, 666, L15
 \bibitem[2024]{xyz}
  Yan, Y.~T., Henkel, C., Menten, K.~M., et al. 2024, A\&A, 686, A205
 \bibitem[2022]{xyz}
  Zhang, Y.-K., Chen, X., Sobolev, A.M., et al. 2022, ApJS, 260, 34
 \bibitem[2023]{xyz}
  Zhang, Y.-K., Chen, X., Song, S.-M., \& Wang, Y.-X. 2023, AJ, 166, 21
 \bibitem[2024]{xyz}
  Zhang, Y.-K., Chen, X., Wang, Y.-X., et al. 2024, AJ, 168, 10
\end{thebibliography}
\end{document}